\documentclass[12pt]{iopart2}
\usepackage{iopams}
\usepackage{epsfig}
\newcommand{\be}{\begin{equation}}
\newcommand{\ee}{\end{equation}}
\newcommand{\bd}{\begin{displaymath}}
\newcommand{\ed}{\end{displaymath}}
\newcommand{\BE}{\begin{eqnarray}}
\newcommand{\EE}{\end{eqnarray}}

\newcommand{\bra}{\left\langle}
\newcommand{\ket}{\right\rangle}

\newcommand{\sgn}{{\rm sgn}}

\newcommand{\id}{{\rm 1\!\!I}}

\newcommand{\bC}{\ensuremath{\mathbf{C}}}

\newcommand{\bG}{\ensuremath{\mathbf{G}}}

\newcommand{\bR}{\ensuremath{\mathbf{R}}}

\newcommand{\boldvarphi}{{\mbox{\boldmath $\varphi$}}}

\newcommand{\tp}{t^\prime}

\newcommand{\chip}{\chi^\prime}
\newcommand{\etabar}{\overline{\eta}}

\begin{document}

\title[Stationary states of a spherical Minority Game with ergodicity breaking]{Stationary states of a spherical Minority Game with ergodicity breaking}

\author{Tobias Galla\dag\ddag, David Sherrington\ddag}
%\footnote[3]{To
%whom correspondence should be addressed (romneya.robertson@iop.org)}
%}

\address{\dag\ International Center for Theoretical Physics, Strada Costiera 11, 34014 Trieste, Italy and CNR-INFM, Trieste-SISSA Unit, V. Beirut 2-4, 34014 Trieste, Italy}
\address{\ddag\ The Rudolf Peierls Centre for Theoretical Physics,
Department of Physics, University of Oxford, 1 Keble Road, Oxford OX1
3NP, UK}

\begin{abstract}
Using generating functional and replica techniques, respectively, we
study the dynamics and statics of a spherical Minority Game (MG),
which in contrast with a spherical MG previously presented in
\cite{GallCoolSher03} displays a phase with broken ergodicity and
dependence of the macroscopic stationary state on initial
conditions. The model thus bears more similarity with the original
MG. Still, all order parameters including the volatility can computed
in the ergodic phases without making any approximations. We also
study the effects of market impact correction on the phase
diagram. Finally we discuss a continuous-time version of the model as
well as the differences between on-line and batch update rules. Our
analytical results are confirmed convincingly by comparison with
numerical simulations. In an appendix we extend the analysis of
\cite{GallCoolSher03} to a model with general time-step, and compare
the dynamics and statics of the two spherical models.
\end{abstract}

\pacs{02.50.Le, 87.23.Ge, 05.70.Ln, 64.60.Ht}

\ead{\tt galla@ictp.trieste.it, sherr@thphys.ox.ac.uk }

\section{Introduction}
The introduction of continuous versions of spin systems with discrete
degrees of freedom is common in statistical mechanics and has led to a
better analytical understanding of many basic features displayed by
various spin models. The appeal of continuous models lies in the fact
that calculations are simplified considerably compared to their
counterparts with discrete Ising spins, so that in many cases a full
analytical solution of the models with continuous spins becomes
possible. The general procedure is to replace the discrete Ising-like
variables $s_i\in\{-1,1\}$ of a given spin system by continuous
variables $\phi_i\in\mathbb{R}$. If we assume that the system is
composed of $N$ such spins ($i=1,\dots,N$), then often a so-called
spherical constraint is imposed on all possible configurations
contributing to the partition function: $\sum_i \phi_i^2=N$. While the
partition function of a system with Ising spins is a sum over
configurations corresponding to the corners of an $N$-dimensional
hypercube, this sum is effectively replaced by an integral over the
surface of the sphere running through these corners. The hope is that
the simplified, spherical model still captures the relevant physical
features of the original model and allows to shed some light on its
basic phenomenology despite the simplification.

The first example of a spherical model is the one of a ferromagnet
introduced more than fifty years ago by Berlin and Kac
\cite{BerlinKac}. It is the continuous analogue of the Ising model,
and can be solved exactly in all dimensions, and exhibits a phase
transition corresponding to the onset of spontaneous
magnetization. Spherical counterparts of the Sherrington-Kirkpatrick
model have played and are playing an important role in the study of
spin-glasses, and similar mean-field models with higher-order
couplings between the continuous spins are able to capture some of the
essential features of structural glasses. Recently a spherical version
of the Hopfield model of neural networks has been considered in
\cite{sphericalhopfield}, and a spherical Minority Game (MG) was
introduced in \cite{GallCoolSher03}. The present work is a
continuation of \cite{GallCoolSher03}, and we will here devise an
alternative way of formulating a spherical limit of the MG.

The Minority Game \cite{ChalZhan97} is a simplified model of
interacting traders in a financial market and has become one of the
most studied models in the econophysics community
\cite{Book1,Book2,Book3}. From the point of view of statistical mechanics the MG is a fully connected mean-field disordered system. Analytical progress can therefore be made using the standard tools of equilibrium and non-equilibrium statistical mechanics, such as replica theory and generating functionals. With these methods an ergodic/non-ergodic phase transition has been identified in the MG, and static order parameters in the ergodic regime computed within a replica-symmetric ansatz agree with numerical simulations. The dynamical approach furthermore allows one to formulate a self-consistent theory  for all regimes of the game in terms of a single effective-agent process \cite{Book2}. This effective problem is fully equivalent to the coupled multi-agent dynamics of the MG in the thermodynamic limit and analytical results for persistent order parameters in the ergodic phase can be obtained from this effective process in agreement with the results derived from the statics. These solutions are now regarded as exact. A proper understanding of the non-ergodic phase is however still awaited. Moreover, for the conventional MG the so-called `market volatility' (one of the MG's main observables) cannot be computed exactly and results are here restricted to approximations even in the ergodic phase.
 
In \cite{GallCoolSher03} a spherical version of the MG was presented,
offering the possibility of a more complete solution. The non-linear
dynamics of the conventional MG was replaced by effectively linear
equations, resulting in an exactly solvable harmonic model. Despite
the simplicity of the update rules this model was shown to exhibit a
non-trivial phase diagram with three distinct phases and different
types of phase transitions.  All stationary observables (including the
volatility) can here be computed in all phases without making any
approximations. The model discussed in
\cite{GallCoolSher03} however has only limited similarity compared to
the original MG. For example the behaviour of the volatility turns out
to be quite different in both models, and crucially the spherical
model presented in \cite{GallCoolSher03} does not exhibit any phases
in which the macroscopic dynamics and the resulting stationary states
depend on initial conditions. This is in sharp contrast with the
original MG, in which initial conditions have been found to be
relevant in the non-ergodic phase and where the resulting memory
effects have to a great part been responsible for the interest of
statistical physicists in the MG.

In this paper we will therefore consider a spherical limit of the MG,
which is different from the route taken in \cite{GallCoolSher03},
leading to a model with a phenomenology closer to that of the original
MG. In particular the model studied in this paper displays
non-ergodicity and memory effects. Nevertheless it is exactly
solvable, and we are able to derive expressions for the volatility in
the ergodic regime without making any approximations. We here choose
to concentrate on the stationary states of the model and do not
address its transient dynamics. While we present some numerical
results for the non-ergodic phase a more detailed analysis would here
require a full solution for the transient dynamics in order to keep
track of the effects of the state from which the dynamics is
started. This is beyond the scope of the present paper but might be
considered in later studies.

The remainder of this paper is organized as follows: In section
\ref{sec:intro} we introduce the model, section \ref{sec:gf} is
concerned with the study of the dynamics, and in section
\ref{sec:phasediagram} we discuss the resulting phase diagram. The
statics are considered in section \ref{sec:replica}. Finally we
present a comparison of the batch model with its continuous time limit
(section \ref{sec:conttime}) and with its on-line counterpart (section
\ref{sec:online}), before we draw conclusions in the final
section. The appendix provides a discussion of some aspects of the
earlier spherical MG not addressed in \cite{GallCoolSher03} as well as
a comparison of the two spherical models.

\section{Model Definitions}\label{sec:intro}
Before turning to spherical versions of the MG, it will be useful to
briefly summarise the definition of the MG in its conventional
setup. The MG describes an ensemble of $N$ agents, whom we shall label
by Roman indices. At each round $t$ of the game, each agent $i$ takes
a binary trading decision $b_i(t)\in\{-1,1\}$ in response to the
observation of a publicly available piece of information
$I_{\mu(t)}$. The aim of each agent is then to make the opposite
choice to the majority of agents. In this paper we will assume that
$\mu(t)$ is chosen randomly and independently from a flat distribution
over the set $\mu(t)\in\{1,\dots,\alpha N\}$ at each time-step. This
is the so-called MG with random external information. The key control
parameter of the model is the ratio $\alpha$ of the number of possible
values of the information over the number of players. We will only
consider the case in which $\alpha$ does not scale with $N$,
i.e. $\alpha={\cal O}(N^0)$. The (re-scaled) total market at time-step
$t$ is then defined as $A(t)=N^{-1/2}\sum_i b_i(t)$. To take trading
decisions each agent $i$ has two strategies
$\bR_{ia}=(R_{ia}^1,\dots,R_{ia}^{\alpha N})\in\{-1,1\}^{\alpha N}$ at
his or her disposal, with $a=\pm 1$\footnote{The MG problem can be generalised to include more strategies per agent, but here we concentrate on this technically simpler but still representative case.}. These effectively act as look-up
tables mapping the values $\mu$ of the external information onto a
trading decision $R_{ia}^\mu$. If agent $i$ decides to use strategy
$s_i(t)\in\{-1,1\}$ in round $t$ his or her bid at this stage will be
$R_{is_i(t)}^{\mu(t)}$. In this paper we will assume that all
strategies are drawn randomly and independently before the start of
the game and that the components $\{R_{ia}^\mu\}$ take the values $\pm
1$ with equal probability; the strategies then remain fixed, they
represent the quenched disorder of the problem.  In order to decide
which of their two strategies to use each agent keeps track of the
performance of both of his or her strategies by assigning virtual
scores $p_{ia}(t)$ based on what would have happened had he or she
always played that particular strategy. These scores are updated
according to
\be
p_{ia}(t+1)=p_{ia}(t)-\frac{R_{ia}^{\mu(t)}}{\sqrt{N}}\left[A(t)-\frac{\kappa}{\sqrt{N}}(R_{is_i(t)}^{\mu(t)}-R_{ia}^{\mu(t)})\right].
\ee
Note that the minus sign in front of the square bracket ensures that
strategies which would have produced a minority decision are
rewarded. The term proportional to $\kappa$ takes into account a
correction of the impact of trader $i$'s action on the total market
bit, and can be seen as an analogue of an Onsager reaction term; see
\cite{DeMaMa01,HeimDeMa01} for details. The additional model parameter $\kappa$
can take values $0\leq\kappa\leq 1$, with $\kappa=0$ corresponding to
the absence of impact-correction and $\kappa=1$ to so-called
`sophisticated agents' or `full impact correction'\footnote{In the
conventional MG the impact-correction term was found to have crucial
consequences for the properties of the model. The game without impact
correction shows an ergodic/non-ergodic phase transition, but no
replica-symmetry breaking. In the game with impact correction
($\kappa>0$) a de Almeida-Thouless transition of a different type,
along with the onset of long-term memory at finite integrated
response, is found \cite{DeMaMa01,HeimDeMa01}.}. In the conventional MG each
player then uses the strategy with the highest score,
i.e. $s_i(t)=\mbox{arg max}_a\, p_{ia}(t)$. In order to determine
which of the two strategies to use, it is sufficient to consider the
score difference $q_i(t)=\frac{1}{2}[p_{i,1}(t)-p_{i,-1}(t)]$, then
one has $s_i(t)=\sgn[q_i(t)]$, and upon introduction of
$\bomega_i=\frac{1}{2}[\bR_{i,1}+\bR_{i,-1}]$ and
$\bxi_i=\frac{1}{2}[\bR_{i,1}-\bR_{i,-1}]$ the bid of player $i$ at
time $t$ may be written as
$b_i(t)=\omega^{\mu(t)}+s_i(t)\xi_i^{\mu(t)}$. The evolution of the
$\{q_i\}$ is then given by
\be
q_i(t+1)=q_i(t)-\frac{\xi_i^{\mu(t)}}{\sqrt{N}}\bigg[\Omega^{\mu(t)}+N^{-1/2}\sum_j\xi_j^{\mu(t)}s_j(t)-\frac{\kappa}{\sqrt{N}}\xi_i^{\mu(t)} s_i(t)\bigg],
\ee
where $\bOmega=N^{-1/2}\sum_j\bomega_j$. This defines the so-called `on-line' MG. Alternatively one may consider models in which the agents update the $\{q_i(t)\}$ only every ${\cal O}(N)$ time-steps. This leads to an effective average over all possible values of the external information and results in the so-called batch MG \cite{HeimCool01}:
\be\label{eq:convbatch}
q_i(t+1)=q_i(t)-h_i-\sum_j J_{ij}s_j(t)+\kappa\alpha s_i(t).
\ee
Here $J_{ij}=2N^{-1}\bxi_i\cdot\bxi_j$ and  $h_i=2N^{-1}\bxi_i\cdot\bOmega$. 

These batch and on-line  update rules provide suitable starting points for the introduction of spherical versions of the MG. In \cite{GallCoolSher03} Eq. (\ref{eq:convbatch}) was replaced by (only the case $\kappa=0$ was considered there)
\be\label{eq:oldspherical0}
[1+\lambda(t+1)]q_i(t+1)=q_i(t)-h_i-\sum_j J_{ij}q_j(t)
\ee
and the $\{\lambda(t)\}$ were chosen to impose a global spherical constraint $\sum_i q_i(t)^2=Nr^2$ for all $t$. Although inspired by the spherical limits sometimes imposed on models of magnetic systems where Ising variables $s_i=\pm 1$ are replaced by continuous spins $\phi_i$ subject only to the same global constraint $\sum_{i=1}^N \phi_i^2=N=\sum_{i=1}^N s_i^2$, without the local constraint $s_i^2=1$, the situation in the earlier spherical MG model is at variance with the original MG in that the latter does not obey $q_i^2=\mbox{constant}$. Rather it is $s_i=\sgn[q_i]$ which obeys $s_i^2=1$ for all $i$ in the MG.
%This is slightly different from the spherical limits usually taken in the context of models of magnetic systems as the $\{q_i\}$ in the conventional MG do not satisfy any spherical constraint, but only the $s_i(t)=\sgn[q_i(t)]$. 
In the present paper we will therefore consider the replacement $s_i(t)\to\phi_i(t)$ with continuous variables $\{\phi_i(t)\}$, and will impose a spherical constraint only on the $\{\phi_i\}$, but not on the $\{q_i\}$. Specifically, we will consider a batch model defined by
\be\label{eq:batch}
q_i(t+1)=q_i(t)-h_i-\sum_j J_{ij}\phi_j(t)+\kappa\alpha \phi_i(t),
\ee
where
\be \label{eq:sphericalconstraint}
\phi_i(t)=\frac{q_i(t)}{\lambda(t)}, \qquad \lambda(t)=\bigg[N^{-1}\sum_i q_i(t)^2\bigg]^{1/2},
\ee
so that the $\{\phi_i(t)\}$ obey the spherical constraint $\sum_i \phi_i(t)^2=N$ for all $t$, in analogy to the $\{s_i(t)=\pm 1\}$ in the original MG. The corresponding on-line model is defined by
\be\label{eq:sphericalonline}
q_i(t+1)=q_i(t)-\frac{\xi_i^{\mu(t)}}{\sqrt{N}}\bigg[\Omega^{\mu(t)}+N^{-1/2}\sum_j\xi_j^{\mu(t)}\phi_j(t)-\frac{\kappa}{\sqrt{N}}\xi_i^{\mu(t)} \phi_i(t)\bigg],
\ee
where we again impose (\ref{eq:sphericalconstraint}). In principle more general dependencies of the $\{\phi_i\}$ on the $\{q_i\}$ could be considered. We here chose the $\phi_i$ to be proportional to the $q_i$ to guarantee the solvability of the resulting harmonic model. Also we assume $\lambda(t)>0$, i.e. $\phi_i(t)$ is taken to have the same sign as $q_i(t)$.

An interpretation of spherical updates rules in terms of the decision making of the agents was given in \cite{GallCoolSher03}: while the agents in conventional MGs have at each time-step to decide which of their two strategies they wish to use, they are playing linear combinations of their strategies in the spherical case. The two spherical limits differ in the way this linear combination is determined. Moreover the models defined by Eqs. (\ref{eq:batch}) and (\ref{eq:sphericalonline}) allow for runway solutions $q_i(t)\to\pm\infty$ (potentially resulting in an asymptotically divergent normalization factor $\lambda(t)$), which are suppressed by the spherical constraint on the $\{q_i(t)\}$ in the earlier spherical model\footnote{Such runaway solutions, in which the $|q_i(t)|$ increase linearly with $t$, are a standard feature of the conventional MG, corresponding to agents described as `frozen' \cite{Book1,Book2}.}.

Let us finally, in this section, introduce one of the key observables in MGs, the so-called volatility $\sigma^2$. It describes the variance of the total re-scaled market bid $A$, and can be defined as the following long-time average:
\be
\sigma^2=\lim_{\tau\to\infty}\tau^{-1}\sum_{t\leq \tau}A(t)^2
\ee
in on-line models. In batch games an additional average over the external information is to be performed and one has
\be
\sigma^2=\lim_{\tau\to\infty}(\alpha N \tau)^{-1}\sum_{t\leq \tau}\sum_{\mu=1}^{\alpha N} \left(A^\mu(t)\right)^2,
\ee
where $A^\mu(t)=\Omega^\mu+N^{-1/2}\sum_i\xi_i^\mu\phi_i(t)$; for further details see also \cite{HeimCool01, GallCoolSher03, Gall05}.  Note that stochastic trading with the $\{\phi_i(t)\}$ taken randomly and independently at any time step $t$ from the spherical surface $\sum_i \phi^2(t)=N$ would result in $\sigma^2=1$. This is referred to as the random trading limit.

\section{Generating functional analysis}\label{sec:gf}
\subsection{Effective single-agent process and macroscopic dynamics}
We will now proceed by studying the dynamics of the present spherical model using generating functionals. This approach is now standard in the context of MGs, and leads to a self-consistent problem for the correlation and response functions formulated in terms of a single effective trader equation. Due to the similarity of the present model with the conventional batch MG and the spherical model presented in \cite{GallCoolSher03} we can obtain the single effective agent equation corresponding to the process (\ref{eq:batch},\ref{eq:sphericalconstraint}) by some minor modifications of the results of \cite{HeimCool01, GallCoolSher03}. One finds 
\be\label{eq:effagent}
q(t+1)=q(t)+\theta(t)-\alpha\sum_{\tp\leq t} (\id+G)^{-1}_{t\tp}\phi(\tp)+\alpha\kappa\phi(t)+\sqrt{\alpha}\eta(t),
\ee
where $\phi(t)=\frac{q(t)}{\lambda(t)}$. \\
Here $\theta(t)$ is an external perturbation field introduced to generate response functions and $\eta(t)$ is a zero-average Gaussian noise with temporal correlations characterized by the following covariance matrix (with $D_{t\tp}=1+C_{t\tp}$ for all $t,\tp$, and $\id$ the identity matrix):
\be\label{eq:covariance}
\Gamma_{t\tp}=\bra \eta(t)\eta(\tp)\ket_*=[(\id+G)^{-1}D(\id+G^T)^{-1}]_{t\tp}.
\ee
The matrices $\bC$ and $\bG$ and the normalization factors $\blambda=\{\lambda(t)\}$ are the dynamical order parameters of the problem, to be determined self-consistently upon solving
\be\label{eq:scconstraints}
C_{t\tp}=\bra \phi(t)\phi(\tp)\ket_*, \quad G_{t\tp}=\frac{\partial}{\partial\theta(\tp)}\bra \phi(t)\ket_*, \quad C_{tt}=1\qquad \forall t,\tp.
\ee
The brackets $\bra\dots\ket_*$ in (\ref{eq:covariance}, \ref{eq:scconstraints}) refer to averages over realizations of the process (\ref{eq:effagent}), i.e. over the noise $\{\eta(t)\}$.  The resulting self-consistent problem for $\{\bC,\bG,\blambda\}$ is equivalent to the original batch process in the thermodynamic limit $N\to\infty$. In particular, the physical meaning of the matrices $\bC$ and $\bG$ is given by
\BE
C_{t\tp}&=&\lim_{N\to\infty}N^{-1}\sum_i \overline{\bra \phi_i(t)\phi_i{\tp}\ket}, \\
G_{t\tp}&=&\lim_{N\to\infty}N^{-1}\sum_i \frac{\partial}{\partial\theta(\tp)} \overline{\bra \phi_i(t)\ket},
\EE
where $\overline{\cdots}$ denotes an average over the disorder, i.e. over the space of all strategy assignments, and $\bra\dots\ket$ stands for an average over possibly random initial conditions (and/or decision noise in case stochastic trading is considered \cite{CoolHeimSher01}). As usual the single effective agent process is non-Markovian and contains coloured Gaussian noise.

Similar to \cite{GallCoolSher03} it is possible to convert the system (\ref{eq:effagent}, \ref{eq:covariance}, \ref{eq:scconstraints}) into a pair of explicit iterative equations for $\bC$ and $\bG$ and $\blambda$; note that due to the non-linearity in $s_i(t)=\sgn[q_i(t)]$ such a conversion is in general not possible for conventional MGs. We find
\BE
\lambda(t+1)C_{t+1,\tp}-(\lambda(t)+\alpha\kappa)C_{t\tp}&=&\alpha[(\id+G)^{-1}D(\id+G^T)^{-1}G^T]_{t\tp}\nonumber\\
&&-\alpha[(\id+G)^{-1}C]_{t\tp},\label{eq:C} \\
\lambda(t+1)G_{t+1,\tp}-(\lambda(t)+\alpha\kappa)G_{t\tp}&=&-\alpha[(\id+G)^{-1}G]_{t\tp}+\delta_{t\tp}\label{eq:G}.
\EE
Eqs. (\ref{eq:C}, \ref{eq:G}) have to be solved subject to the constraint $C_{tt}=1$ for all $t\geq 0$. Furthermore one finds as in \cite{HeimCool01} for $N\to\infty$ that the rescaled disorder-averaged total bid $\overline{\bra A(t)\ket}$ is zero for all times, and that the disorder-averaged volatility is asymptotically given by the diagonal elements of the covariance matrix of the single-agent noise:
\be\label{eq:voldef0}
\sigma^2=\frac{1}{2}\lim_{\tau\to\infty}\tau^{-1}\sum_{t\leq\tau}[(\id+G)^{-1}D(\id+G^T)^{-1}]_{tt}.
\ee
Already at this stage one may notice that  Eqs. (\ref{eq:C},\ref{eq:G}) allow in principle for non-ergodic behaviour. The value $\lambda(t=0)=\left[\lim_{N\to\infty}N^{-1}\sum_{i=1}^N \overline{\bra q_i(0)^2\ket}\right]^{1/2}$ is not fixed by the dynamics, but is set by the initial conditions from which the dynamics is started. The type of stationary state reached asymptotically may hence depend on the starting point, similarly to what is observed in conventional MGs, and at variance with the behaviour of the spherical model defined by (\ref{eq:oldspherical0}), where the macroscopic dynamics at all times is fully fixed by the model parameter $r$.
%\footnote{In \cite{GallCoolSher03} it was assumed that the initial conditions fulfill $N^{-1}\sum_{i=1}^N q_i(0)^2=r^2$, so that $C_{tt}=r^2$ also for $t=0$. In principle one might there also consider initial biases with a second moment different from $r^2$, resulting in a model with $C_{tt}=r^2$ for all $t>0$, but with $C_{00}\neq r^2$.}. 
We will discuss the effects of different starts $|q_i(0)|=q_0$ for all $i$ with varying bias $q_0$ below.

\subsection{Time-translation invariant ergodic states}
We will consider the system long after any initial equilibration and
focus on time-translation invariant (TTI) solutions of the coupled equations
(\ref{eq:C}, \ref{eq:G}) of the form
\be\label{eq:tti}
\lim_{t\to\infty}C_{t+\tau,t}=C(\tau), \qquad \lim_{t\to\infty}G_{t+\tau,t}=G(\tau).
\ee
In addition we make the standard assumptions of finite integrated response and weak long-term memory:
\be\label{eq:chifinitewltm}
 \chi\equiv\lim_{t\to\infty} \sum_{\tau\leq t} G(\tau)<\infty, \qquad \lim_{t\to\infty} G_{t,\tp}=0\quad\forall \tp \mbox{ finite}.
\ee
The order parameter $\blambda$ comes out as a function of the form
$\lambda(t)=\lambda_0+\lambda_1 t$ asymptotically in simulations, with
$\lambda_0$ and $\lambda_1$ constants. As it turns out we will find
two distinct phases with the above properties:
\begin{itemize}
\item[(i)] an oscillatory phase in which $\lim_{t\to\infty}\lambda(t)$ remains finite, i.e. $\lambda_1=0$ and 
\item[(ii)] a frozen phase in which $\lambda(t)$ grows linearly in time asymptotically, i.e. $\lambda_1>0$. 
\end{itemize}
We will refer to these phases as `bounded' and `unbounded' in the following, and will discuss them separately\footnote{One may argue whether a phase with a linearly growing order parameter $\lambda(t)$ qualifies as time-translation invariant. Note however that the stationary ergodic states of the standard MG also contain runaway solutions of the type $q_i(t)\to\pm\infty$, so-called frozen agents, so that $N^{-1}\sum_i q_i(t)^2$ diverges in time. In the unbounded phase we find that $\lambda(t)/t$ is the stationary quantity asymptotically.}.

\subsubsection{Bounded oscillatory states $(\lambda_1=0)$}\label{sec:o}
We here consider $\lambda(t)\equiv \lambda_0$. Given our TTI ansatz
(\ref{eq:tti}) all matrices in (\ref{eq:C},
\ref{eq:G}) become Toeplitz matrices in the stationary state, and hence they commute. It is then furthermore convenient to introduce Fourier transforms of the
correlation and response functions according to
\be
C(\tau)=\int_{-\pi}^{\pi}\frac{d\omega}{2\pi}e^{i\omega\tau}\widetilde C(\omega),\quad \widetilde C(\omega)=\sum_\tau e^{-i\omega\tau}C(\tau)
\ee
and similarly for $G$. Now upon setting $\lambda(t)\equiv\lambda_0$ in the stationary state, Eqs.  (\ref{eq:C}, \ref{eq:G}) translate into
\BE
\left[\lambda_0(e^{i\omega}-1)-\alpha\kappa\right]\widetilde C(\omega)&=&\frac{\alpha\widetilde D(\omega)\widetilde G(\omega)^*}{|1+\widetilde G(\omega)|^2}-\frac{\alpha\widetilde C(\omega)}{1+\widetilde G(\omega)},\label{eq:Cf} \\ 
\left[\lambda_0(e^{i\omega}-1)-\alpha\kappa\right]\widetilde G(\omega)&=&1-\frac{\alpha \widetilde G(\omega)}{1+\widetilde G(\omega)}\label{eq:Gf}.
\EE
Here $\widetilde G(\omega)^*$ denotes the complex conjugate of $\widetilde
G(\omega)$. Using $\widetilde D(\omega)=\widetilde
C(\omega)+2\pi\delta(\omega)$ as well as the definition of the
integrated response $\chi=\sum_\tau G(\tau)=\widetilde G(0)$ we may write
(\ref{eq:Cf}) as
\be
\bigg[\left\{\lambda_0(e^{i\omega}-1)-\alpha\kappa\right\}|1+\widetilde G(\omega)|^2+\alpha\bigg]\widetilde C(\omega)=2\pi\alpha\chi\delta(\omega).
\ee
From this equation it follows immediately that $\widetilde C(\omega)$ can be non-zero for any $\omega\neq 0$ only if $e^{i\omega}$ is real, i.e. for $\omega=\pi$. Consequently $\widetilde C(\omega)$ must be of the form $\widetilde C(\omega)=2\pi[c_0\delta(\omega)+c_1\delta(\omega-\pi)]$. Equivalently we may write
\be
C(\tau)=c_0+c_1(-1)^\tau.
\ee
Note that we always have $c_0+c_1=1$ due to the spherical constraint $C(\tau=0)=1$. Inserting this ansatz into (\ref{eq:Cf},\ref{eq:Gf}) and setting $\omega=0$ then leads to the following relations
\be\label{eq:eqforomegazero}
\alpha\kappa c_0 = - \frac{\alpha(1+c_0)\chi}{(1+\chi)^2}+\alpha\frac{c_0}{1+\chi}, \qquad \alpha\kappa\chi=-\frac{1+(1-\alpha)\chi}{1+\chi}.
\ee
After some algebra the physically relevant solutions are identified as
\BE 
c_0&=&\left[\left(\frac{1+(1-\kappa)\alpha+\sqrt{[1+(1-\kappa)\alpha]^2-4\alpha}}{2\sqrt{\alpha}}\right)^2-1\right]^{-1},\label{eq:c0osc}\\
\chi&=&[\sqrt{\alpha}\sqrt{1+1/c_0}-1]^{-1}.\label{eq:chiosc}
\EE
This determines $c_0$ and $\chi$ as functions of $\alpha$ and $\kappa$, and simplifies to $c_0=\chi=\frac{1}{\alpha-1}$ for $\kappa=0$. The expression for $c_0$ implies that a fully frozen solution ($c_0=1$) can be realised only at the point given by
\be\label{eq:alphac1}
\alpha_{c1}(\kappa)=\frac{5+4\kappa+3\sqrt{1+8\kappa}}{4(1-\kappa)^2}. 
\ee
This describes a line in the $(\alpha,\kappa)$-plane, so that there
can be no frozen TTI phases with finite integrated response and with a
finite value of $\lambda_0=\lim_{t\to\infty}t^{-1}\sum_{\tau\leq
t}\lambda(\tau)$. We can therefore focus on solutions with
$c_0<1$. These are solutions in which oscillations of the correlation
function persist, so that the oscillation amplitude $c_1=1-c_0$ is
strictly positive.  Setting $\omega=\pi$ in (\ref{eq:Cf},\ref{eq:Gf})
gives
\BE
((-2\lambda_0-\alpha\kappa)(1+\chip)^2+\alpha)(1-c_0)&=&0,\\
(-2\lambda_0-\alpha\kappa)\chip&=&1-\frac{\alpha\chip}{1+\chip}\label{eq:helpeq}.
\EE
Here $\chip=\sum_\tau (-1)^\tau G(\tau)=\widetilde G(\pi)$ measures the
response of the system to persistent oscillatory
perturbations. Looking only for oscillatory solutions ($c_1>0$, $c_0<1$) we
must require $(2\lambda_0+\alpha\kappa)(1+\chip)^2=\alpha$ in this regime. Using this
and (\ref{eq:helpeq}) we find the following solutions valid for
$\alpha>\alpha_{c1}(\kappa)$:
\be\label{eq:solosc2}
\lambda_0=\frac{1}{2}[\alpha+1+2\sqrt{\alpha}-\alpha\kappa], \qquad \chip=-\frac{1}{1+\sqrt{\alpha}}.
\ee
The persistent order parameters in the ergodic bounded oscillatory state are thus fully described by Eqs. (\ref{eq:chiosc}, \ref{eq:c0osc}, \ref{eq:solosc2}). The consistency of this ansatz breaks down whenever $\chi\to\infty$ or $c_0\to 1$. From (\ref{eq:alphac1}) one observes that $c_0\to 1$ at a value of $\alpha=\alpha_{c1}\geq 2$, whereas (\ref{eq:chiosc}) dictates that $\chi$ remains finite for all $\alpha\geq 1$ for any fixed value of $\kappa\geq 0$. Hence we conclude that the oscillation amplitude vanishes before anomalous response sets in as $\alpha$ is lowered at a fixed value of $\kappa$. We thus expect a transition to a frozen state at finite integrated response.
 
Finally, using the fact that $C(\tau)=c_0+(1-c_0)(-1)^\tau$ the volatility in the stationary state can be obtained from Eq. (\ref{eq:voldef0}) as
\be\label{eq:voldef}
\sigma^2=\frac{1}{2}\left[(\id+G)^{-1}D(\id+G^T)^{-1}\right](0)=\frac{1}{2}\left[\frac{1+c_0}{(1+\chi)^2}+\frac{1-c_0}{(1+\chip)^2}\right].
\ee
Note that this is an exact result, with no approximations made at any stage.

\subsubsection{Unbounded frozen states $(\lambda_1>0)$}\label{sec:f}

A second type of time-translation invariant solutions can be found
upon making the ansatz
\be
C_{t,\tp}\equiv 1, \quad G_{t,\tp}=G(t-\tp)\qquad \forall t,\tp,
\ee
but giving up the assumption that $\lambda(t)=\lambda_0$ be
constant. Instead, these solutions correspond to a fully frozen state,
and we find in numerical simulations that 
\be
\lambda(t)=\lambda_0+\lambda_1 t
\ee
for a constant $\lambda_1>0$ (yet to be determined) and that the
$q_i(t)$ take the form $q_i(t)=v_i t$.  In order to compute the
persistent order parameters, it here turns out to be convenient to
work with the effective process rather than with the closed equations
for $C$ and $G$. Upon making the ansatz $q(t)=vt$ for solutions of the
single-effective agent process, and assuming that
$\theta(t)\equiv\theta$ is a persistent perturbation one performs an
average of the effective process (\ref{eq:effagent}) over the time $t$
(a similar time-averaging procedure for the effective process of the
conventional batch MG is discussed in \cite{HeimCool01}) and obtains
\be\label{eq:vnull}
v=\theta-\frac{\alpha v}{\lambda_1(1+\chi)}+\frac{\alpha\kappa v}{\lambda_1}+\sqrt{\alpha}\etabar.
\ee
Here $\etabar=\lim_{T\to\infty} T^{-1}\sum_{t\leq T}\eta(t)$ is a static Gaussian variable of zero average and with variance $\bra \etabar^2\ket_{\etabar}=(1+c_0)/(1+\chi)^2$ (with $c_0=1$ in the frozen state under investigation). The only purpose of $\theta$ is to generate response functions, and noting that derivatives with respect to $\theta$ can (up to a factor $\sqrt{\alpha}$) be replaced by derivatives with respect to $\etabar$ we may set $\theta=0$ and write 
\be\label{eq:v}
v=\frac{\sqrt{\alpha}\etabar}{1+\frac{\alpha}{\lambda_1}\left(\frac{1}{1+\chi}-\kappa\right)}.
\ee
Self-consistency then demands
\be
\lambda_1=\sqrt{\bra v^2\ket_{\etabar}}, \qquad \chi=\frac{1}{\lambda_1\sqrt{\alpha}}\bra \frac{\partial v}{\partial \etabar}\ket_{\etabar}.
\ee
Using (\ref{eq:v}) these relations are given by
\be
\lambda_1=\frac{\sqrt{\alpha}}{1+\frac{\alpha}{\lambda_1}\left(\frac{1}{1+\chi}-\kappa\right)}\frac{\sqrt{2}}{(1+\chi)},\quad \chi=\frac{1}{\lambda_1\left(1+\frac{\alpha}{\lambda_1}\left(\frac{1}{1+\chi}-\kappa\right)\right)},
\ee
and the solutions are found to be
\be
\lambda_1=-1-\alpha+\frac{3\sqrt{\alpha}}{\sqrt{2}}+\alpha\kappa, \qquad \chi=\frac{1}{\sqrt{2\alpha}-1}.
\ee
In particular this means that the assumption of finite integrated response $\chi$ breaks down at $\alpha=\alpha_{c2}=1/2$ so that we expect a transition to a non-ergodic phase.  Note that this happens irrespective of the value of $\kappa$, in particular we could not find any signs of memory onset at finite $\chi$ in the case $\kappa>0$, as it is observed for the conventional MG with impact correction \cite{HeimDeMa01}\footnote{We have performed a calculation along the lines of \cite{HeimDeMa01}, adapted to the present model, and could not observe any continuous onset of long-term memory. Also our simulations show near perfect agreement with the ergodic theory for $\kappa>0$ so that we have no reason to suspect the breakdown of the assumption of weak long-term memory.}.

The volatility in the frozen ergodic phase is obtained by inserting the above expression for $\chi$ into (\ref{eq:voldef}) (with $c_0=1$), and one finds
\be\label{eq:volF}
\sigma^2=\frac{1}{2}\bigg[\sqrt{2}-\frac{1}{\sqrt{\alpha}}\bigg]^2.
\ee
This expression is exact and valid in the interval
$\alpha\in[\alpha_{c2}=1/2,\alpha_{c1}(\kappa)$ for any fixed value of
$\kappa$. The zeroes of $\lambda_1=\lambda_1(\alpha)$ are found as:
\be
\alpha=\frac{5+4\kappa\pm 3\sqrt{1+8\kappa}}{4(1-\kappa)^2}. 
\ee
While the upper value agrees with $\alpha_{c1}$ marking the boundary between the frozen and oscillatory phases with finite integrated response, we find that the lower zero of $\lambda_1$ has no relevance for $\kappa>0$ as it occurs at a lower value of $\alpha$ than the onset of anomalous response at $\alpha_{c2}=1/2$.

\begin{figure}[t]
\vspace*{-0mm} \hspace*{5mm} \setlength{\unitlength}{1.0mm}
\begin{tabular}{cc}
\begin{picture}(100,55)
\put(-8,5){\epsfysize=50\unitlength\epsfbox{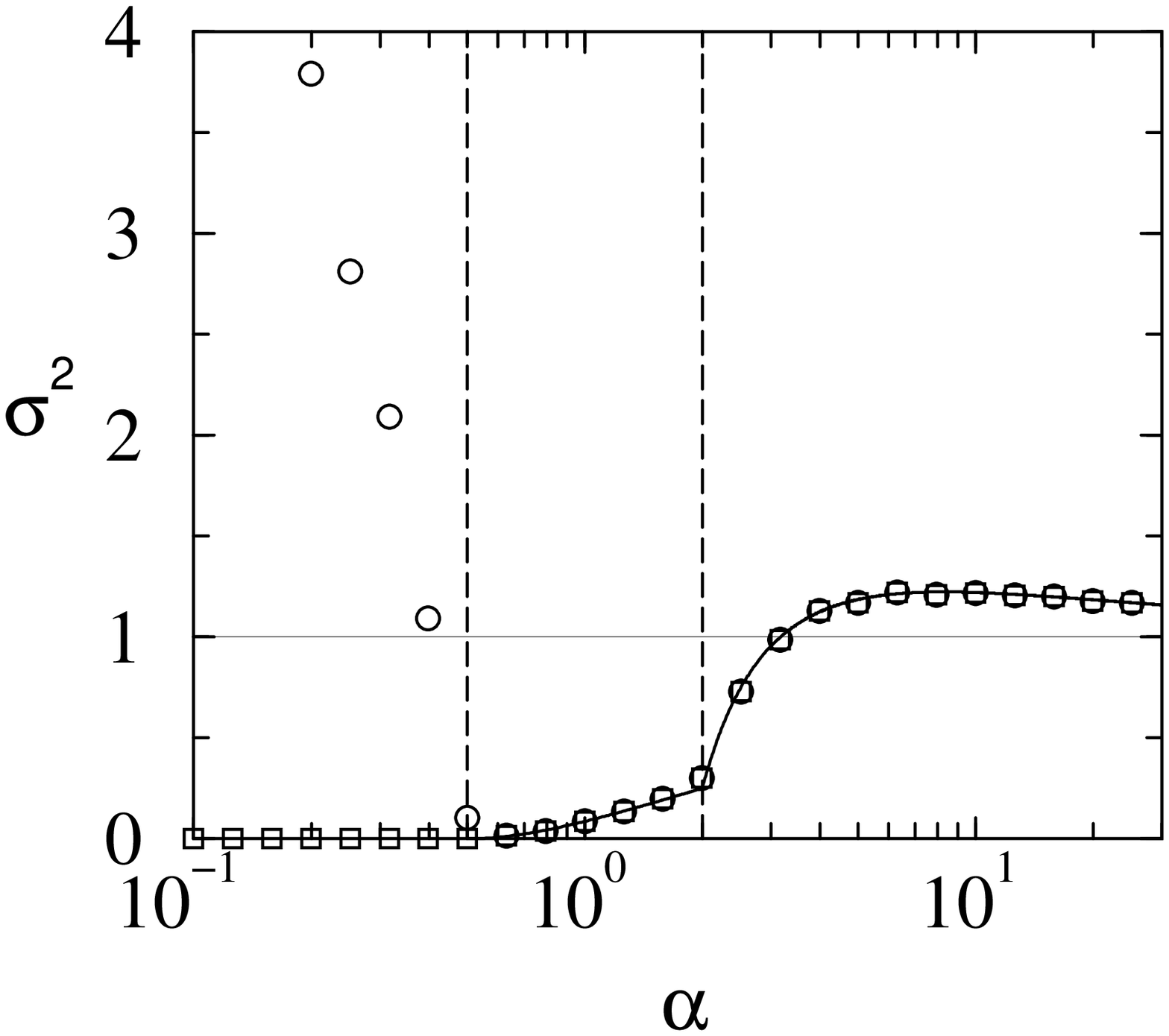}}
\end{picture} ~~~~~~~~ & 
\begin{picture}(100,55)
\put(-43,5){\epsfysize=50\unitlength\epsfbox{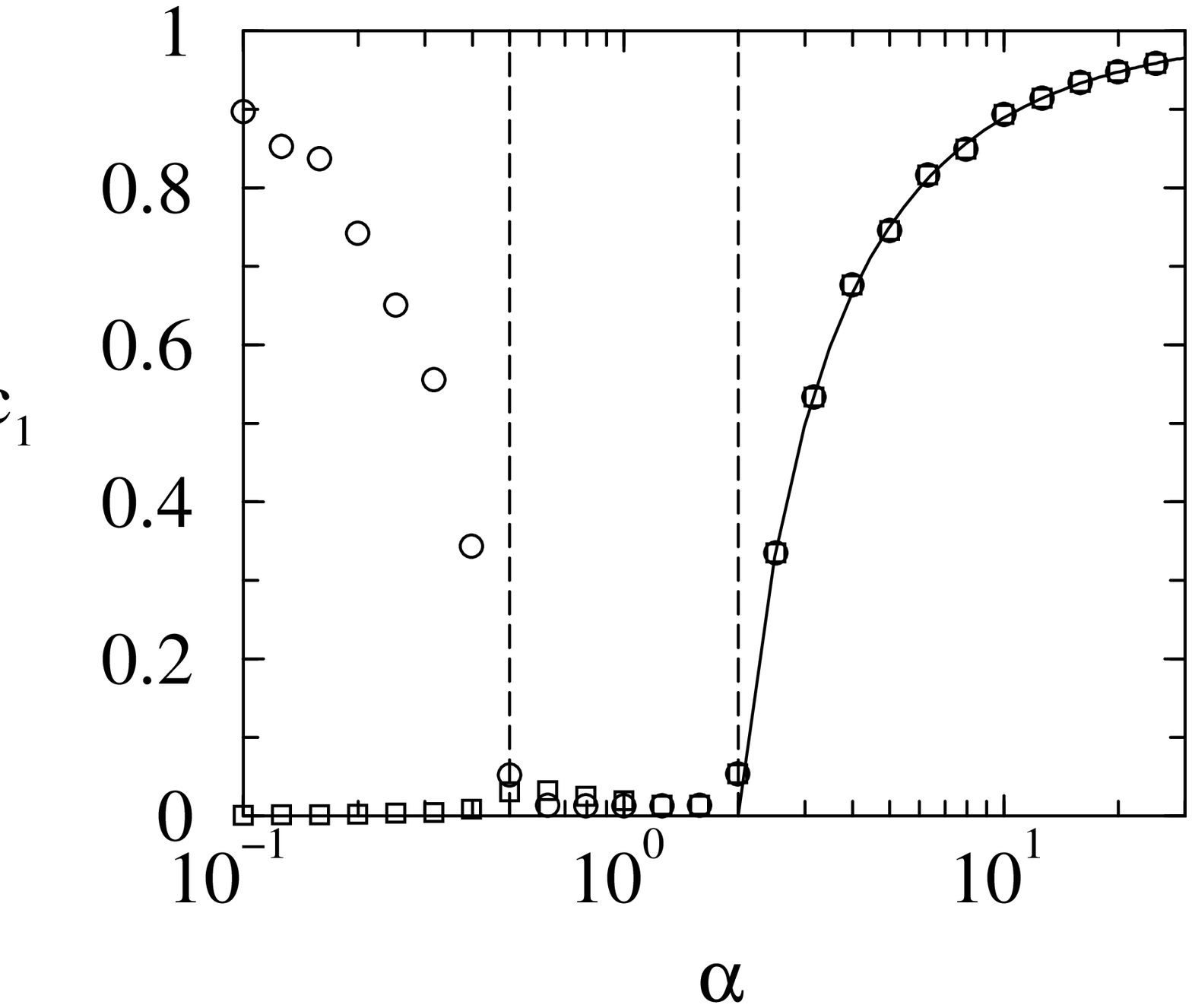}}
\put(-96,49){\small \bf F} 
\put(-112,49){\small \bf AFO}   
\put(-77,49){\small \bf O} 
\put(-16,49){\small \bf F} 
\put(-32,49){\small \bf AFO}   
\put(0,49){\small \bf O} 
\end{picture}
\\
\begin{picture}(100,55)
\put(-8,5){\epsfysize=50\unitlength\epsfbox{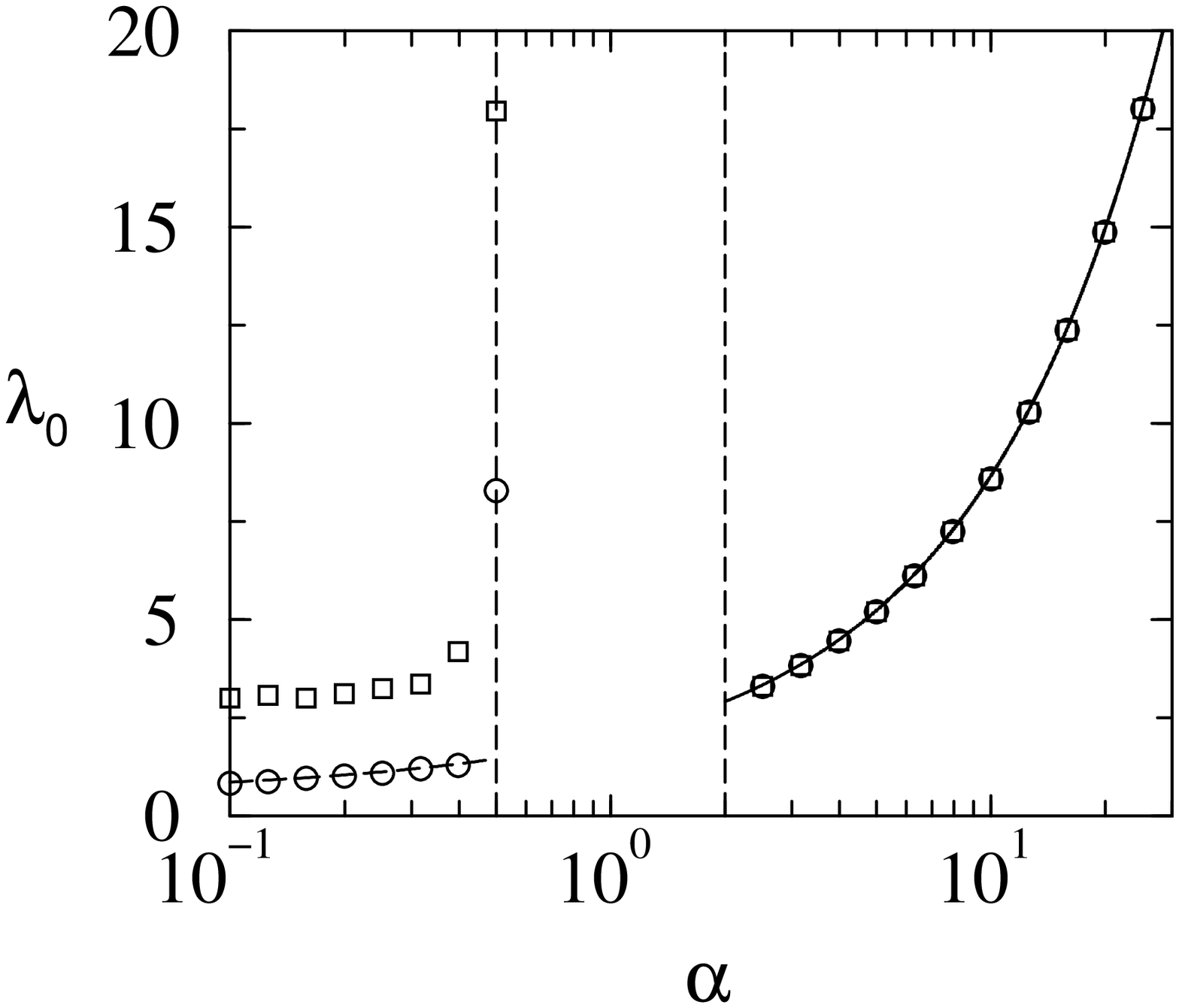}}
\end{picture} ~~~~~~~~ & 
\begin{picture}(100,55)
\put(-43,5){\epsfysize=50\unitlength\epsfbox{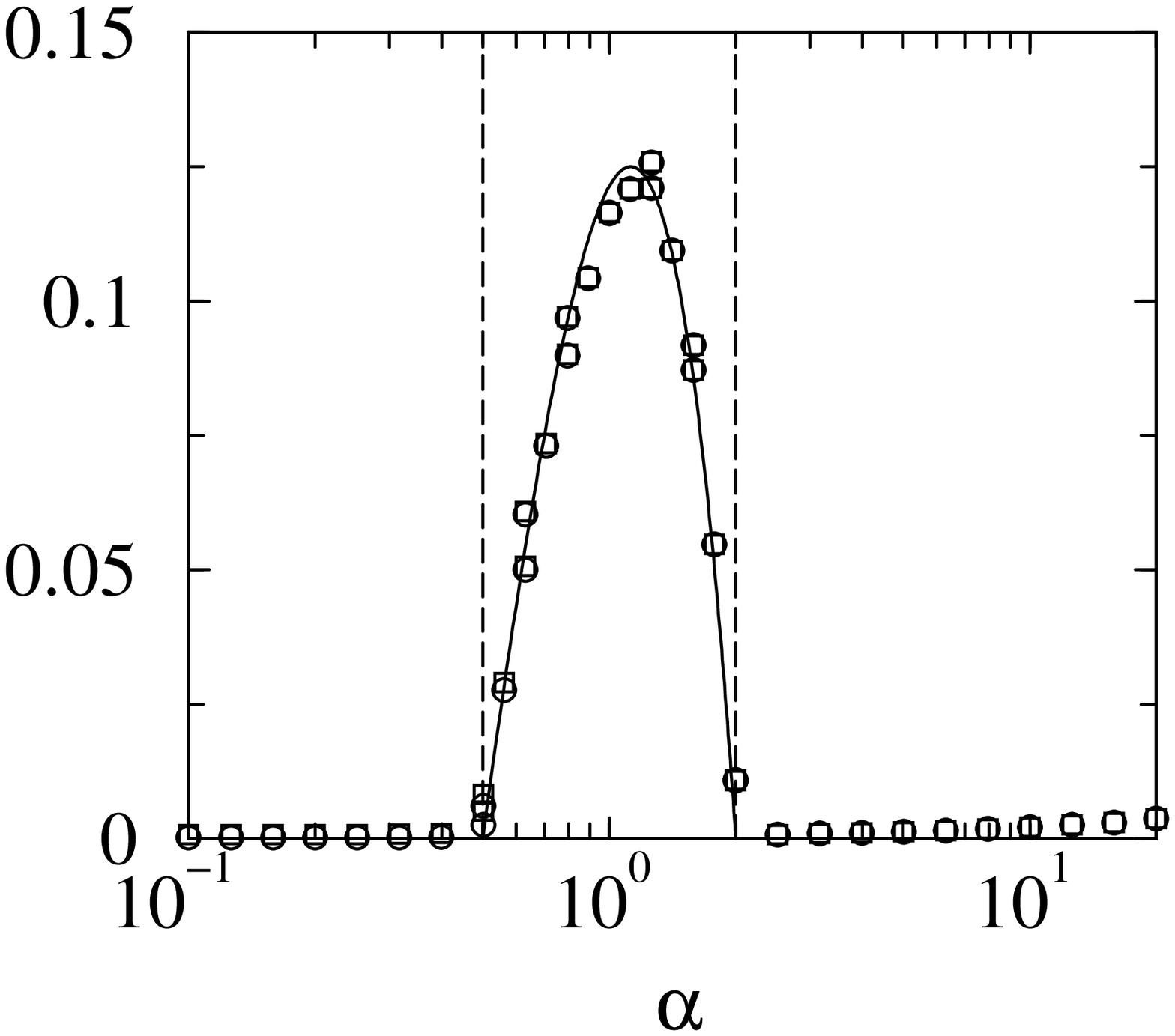}}
\put(-96,49){\small \bf F} 
\put(-112,49){\small \bf AFO}   
\put(-77,49){\small \bf O} 
\put(-12,49){\small \bf F} 
\put(-28,49){\small \bf AFO}   
\put(4,49){\small \bf O}
%\put(-8,29){\Large $r$} \put( 19,3){\Large $\alpha$}
\end{picture}
\end{tabular}
\vspace*{4mm} \caption{Volatility $\sigma^2$, oscillation amplitude $c_1=1-c_0$, and order parameters $\lambda_0$ and $\lambda_1$ for the game without impact correction ($\kappa=0$) for different initial conditions. Markers are from simulations with $q_0=0.01$ (circles) and $q_0=3.0$ (squares) respectively. Solid lines are the theoretical predictions in the ergodic phases (continued as a dashed line into the non-ergodic phase for $\lambda_0$). ${\bf O}$ labels the oscillatory phase of section \ref{sec:o}, ${\bf F}$ the frozen unbounded phase (section \ref{sec:f}) and ${\bf AFO}$ the phase with anomalous response, in which oscillatory and frozen states can be found (section \ref{sec:afo}). The vertical dashed lines mark the locations of the O$\leftrightarrow$F and F$\leftrightarrow$AFO transitions, $\sigma^2=1$ is the random trading limit. Simulations are for $N=500$ agents, run for $2000$ batch steps (additionally some runs were performed for up to $20000$ steps to confirm equilibration in the non-ergodic phase). All data are averages over $10$ realisations of the disorder. \label{fig:etaeq0}}
\end{figure}

\begin{figure}[t]
\vspace*{-0mm} \hspace*{5mm} \setlength{\unitlength}{1.0mm}
\begin{tabular}{cc}
\begin{picture}(100,55)
\put(-8,5){\epsfysize=50\unitlength\epsfbox{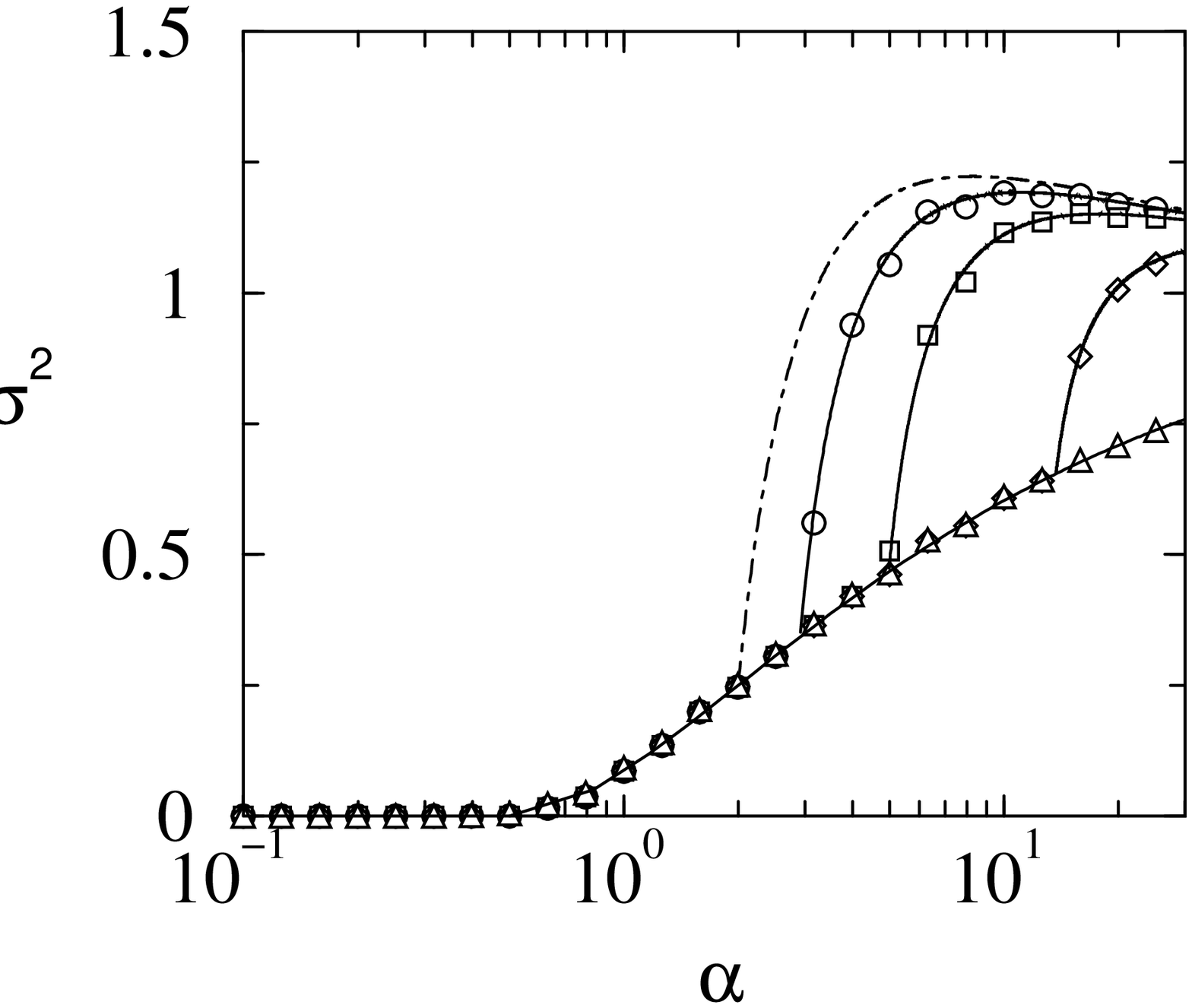}}
\end{picture} ~~~~~~~~ & 
\begin{picture}(100,55)
\put(-43,5){\epsfysize=50\unitlength\epsfbox{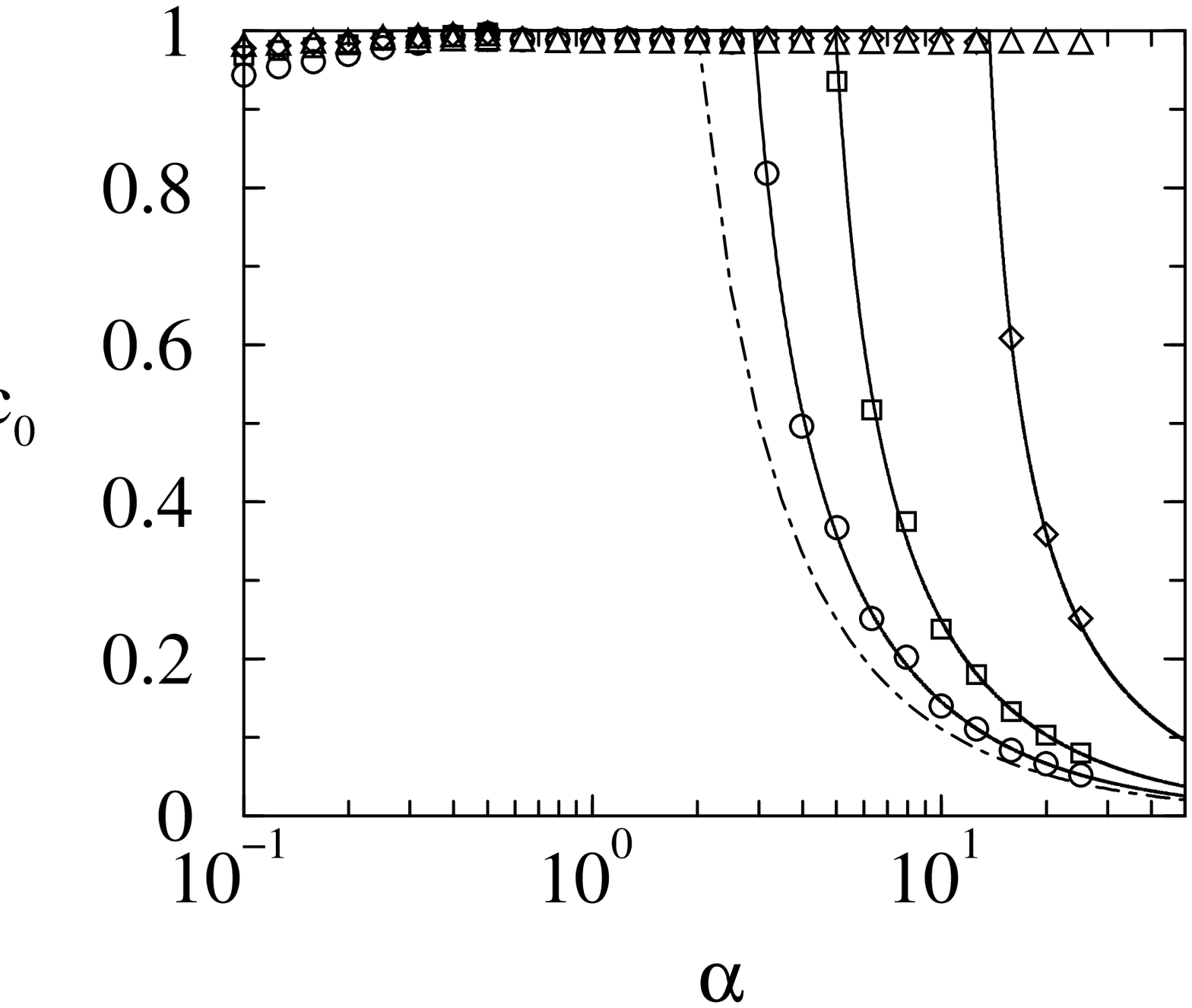}}
\end{picture}
\\
\begin{picture}(100,55)
\put(-8,5){\epsfysize=50\unitlength\epsfbox{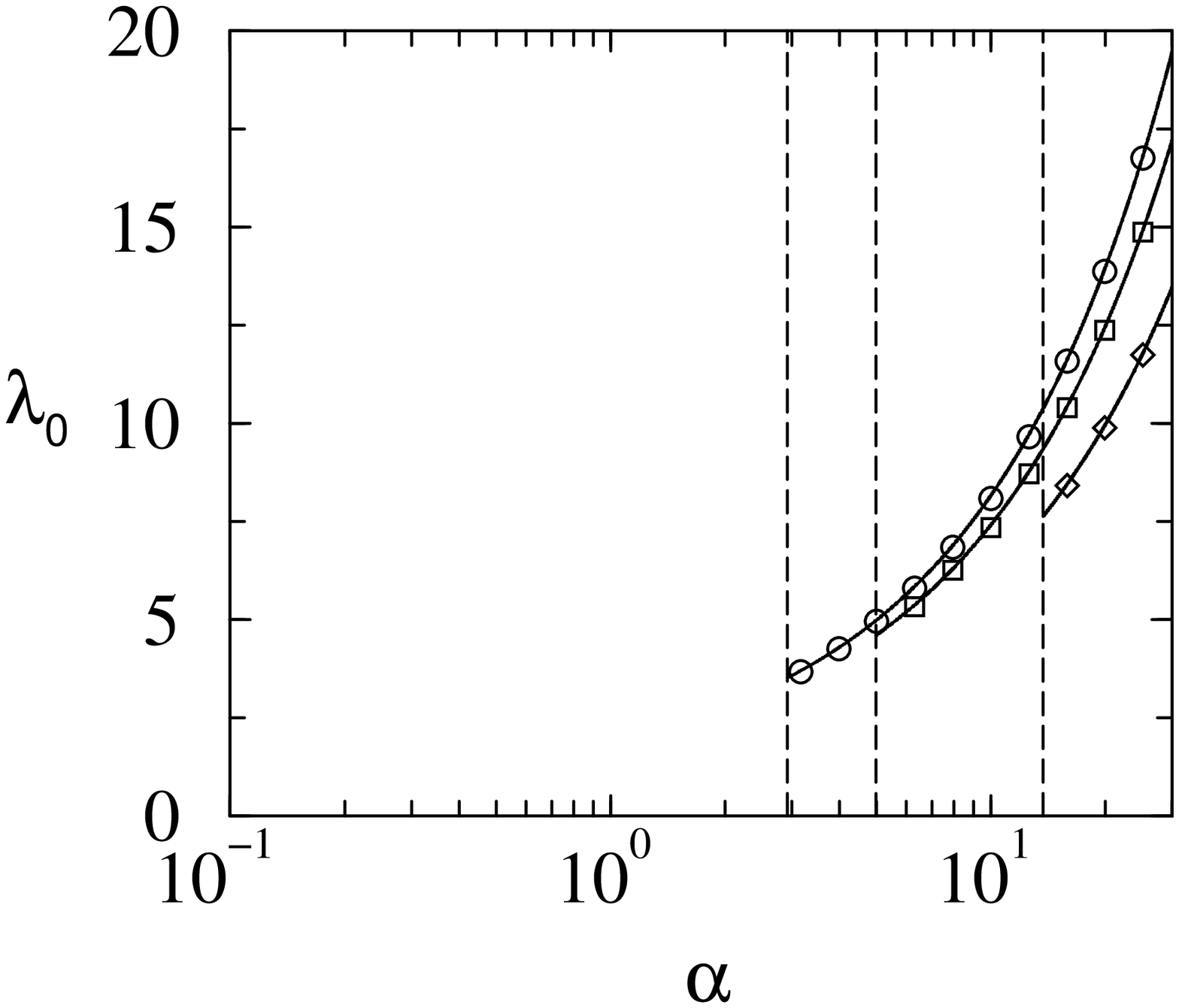}}
\end{picture} ~~~~~~~~ & 
\begin{picture}(100,55)
\put(-43,5){\epsfysize=50\unitlength\epsfbox{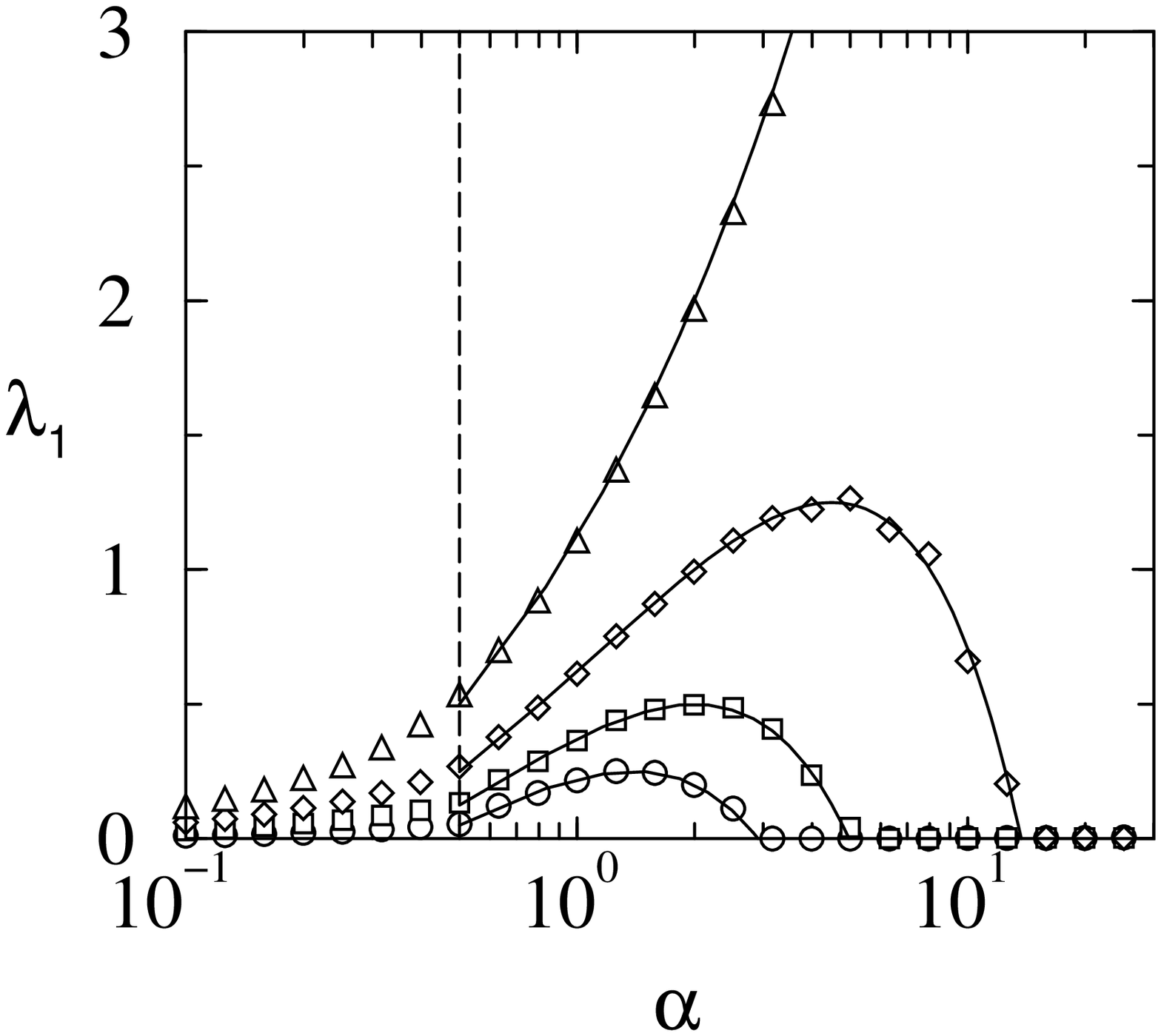}}
\put(-19,2){\vector(0,1){5}}
\put(-22,-2){$\alpha_{c2}$}
\put(-79,2){\vector(0,1){5}}
\put(-83,-2){$\alpha_{c1}(\kappa)$}
\end{picture}
\end{tabular}
\vspace*{4mm}  \caption{Volatility $\sigma^2$, persistent correlation $c_0$, order parameters $\lambda_0$ and $\lambda_1$ for the game with impact correction.  Results are shown for $\kappa=0.1$ (circles), $\kappa=0.25$ (squares), $\kappa=0.5$ (diamonds) and $\kappa=1.0$ (triangles). Symbols are from simulations with $N=500$ agents, $10$ realisations, initial conditions $q_0=0.01$. Simulations are run for $2000$ batch steps ($20000$ steps for the data for $c_0$ at $\alpha<0.5$ in order to minimize artifacts due to incomplete equilibration). Solid lines: theory. We show the results for $\sigma^2$ and $c_0$ for the game at $\kappa=0$ as dot-dashed lines in the upper two panels for comparison.\label{fig:impcorrfig}}
\end{figure}

We have tested our theoretical predictions for the order parameters in
the ergodic states against direct numerical simulations of the batch
process. Results for the game without impact correction ($\kappa=0$)
are displayed in Fig. \ref{fig:etaeq0}, while
Fig. \ref{fig:impcorrfig} is concerned with the game with impact
correction ($\kappa>0$). In both cases we find near perfect agreement
between the analytical theory and simulations, deviations close to the
transitions points are due to finite-size and/or finite running time
artifacts. The resulting phase diagram will be discussed in more
detail below.

\subsection{The non-ergodic phase}\label{sec:afo}
Throughout the non-ergodic phase one has $\chi=\infty$, and the
stationary state depends on the initial conditions from which the
dynamics is started. Hence, computing the asymptotic order parameters
would require a full study of the intermediate transient dynamics to
keep track of the influence of the starting point. This is beyond the
scope of the present paper; we are here only concerned with the
stationary states themselves. In numerical simulations of the
non-ergodic regime of the game {\em without} impact-correction these
states turn out to be (i) bounded and (ii) either frozen or
oscillatory, so that we must assume that the ansatz $\lambda_1=0$ and
$C(\tau)=c_0+(1-c_0)(-1)^\tau$ is still valid below
$\alpha_{c2}=1/2$. Setting $\chi=\infty$ in Eq. (\ref{eq:voldef})
(which strictly speaking has been derived only under the assumptions
of ergodicity and time-translation invariance) leads to
\be\label{blu}
\sigma^2=\frac{1}{2}\frac{1-c_0}{(1+\chip)^2}.
\ee
\begin{figure}[t]
\vspace*{-26mm} \hspace*{35mm} \setlength{\unitlength}{1.8mm}
\begin{picture}(100,55)
\put(-8,-0){\epsfysize=35\unitlength\epsfbox{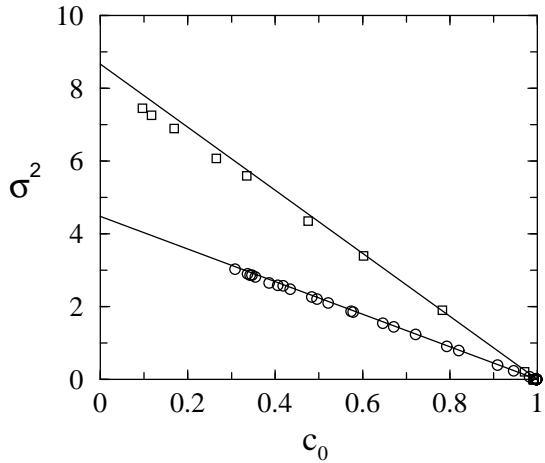}}
\end{picture}
\vspace*{-8mm} \caption{ $\sigma^2$ vs $c_0$ for the game without impact correction ($\kappa=0$) at fixed $\alpha=0.25$ (circles) and $\alpha=0.1$ (squares) respectively. Symbols are from simulations, every marker corresponds to a different starting point $q_i(0)=q_0$ for all $i$ with $q_0$ varying from $0.01$ to $2.0$ in small steps. Leftmost markers correspond to $q_0=0.01$, rightmost symbols to frozen states resulting from high initial bias. Solid lines are Eq. (\ref{eq:sigmasqnonerg}).} \label{fig:volvsc0}
\end{figure}

For the frozen states ($c_0=1$) with diverging integrated response we thus expect the volatility to vanish as confirmed in numerical simulations, see Fig. \ref{fig:etaeq0}.
Furthermore, one realises that in the derivation of (\ref{eq:solosc2}) only the case $\omega=\pi$ in Eqs. (\ref{eq:Cf},\ref{eq:Gf}) is used, so that $\chip$ and $\lambda_0$ can be computed independently of $c_0$ and $\chi$. We may hence conjecture that (\ref{eq:solosc2}) still holds in the non-ergodic regime (at least in approximation).  Note however that  (\ref{eq:solosc2}) can only be expected to be valid in oscillatory states as we make explicit use of the assumption $c_1>0$. Fig. \ref{fig:etaeq0} confirms that $\lambda_0=(\alpha+1+2\sqrt{\alpha})/2$ appears to describe the simulational results also in the non-ergodic regime, as long as the system is initialized at a small bias leading to an oscillatory regime. Insertion into (\ref{blu}) gives
\be\label{eq:sigmasqnonerg}
\sigma^2=\frac{1}{2}\frac{(1+\sqrt{\alpha})^2}{\alpha}(1-c_0).
\ee
As shown in Fig. \ref{fig:volvsc0} this relation turns out to be valid to a very good accuracy. The set of possible solutions in the non-ergodic phase thus appears to be parametrized by $c_0$, with different values attained depending on initial conditions.\footnote{Simulations of the game without impact-correction and at fixed $\alpha<\alpha_{c2}=1/2$ seem to indicate that oscillatory solutions are realised for initial biases $q_0$ smaller than some critical value $q_{0c}(\alpha)$. In this regime $c_0$ is an increasing function of $q_0$. Starting points $q_0>q_{0c}(\alpha)$ lead to frozen states with $c_0=1$ and vanishing volatility. From our simulations  we cannot rule out that $q_{0c}(\alpha)$ may indeed depend on $\alpha$. A similar threshold separating non-ergodic states with diverging and vanishing volatilities respectively  has been identified in \cite{HeimCool01} in the limit $\alpha\to 0$ of the conventional MG.} 

For the game with impact-correction ($\kappa>0$) we could not find any oscillatory solutions below $\alpha=1/2$ in numerical simulations. Independent of initial conditions, all observed states were fully frozen and unbounded in the sense of the above definitions, as indicated by non-zero values of $\lambda_1$ in Fig. \ref{fig:impcorrfig}, and showed a zero volatility for all $\alpha<1/2$. 

\section{Phase diagram and discussion}\label{sec:phasediagram}
\begin{figure}[t]
\vspace*{-26mm} \hspace*{35mm} \setlength{\unitlength}{1.8mm}
\begin{picture}(100,55)
\put(-8,-0){\epsfysize=40\unitlength\epsfbox{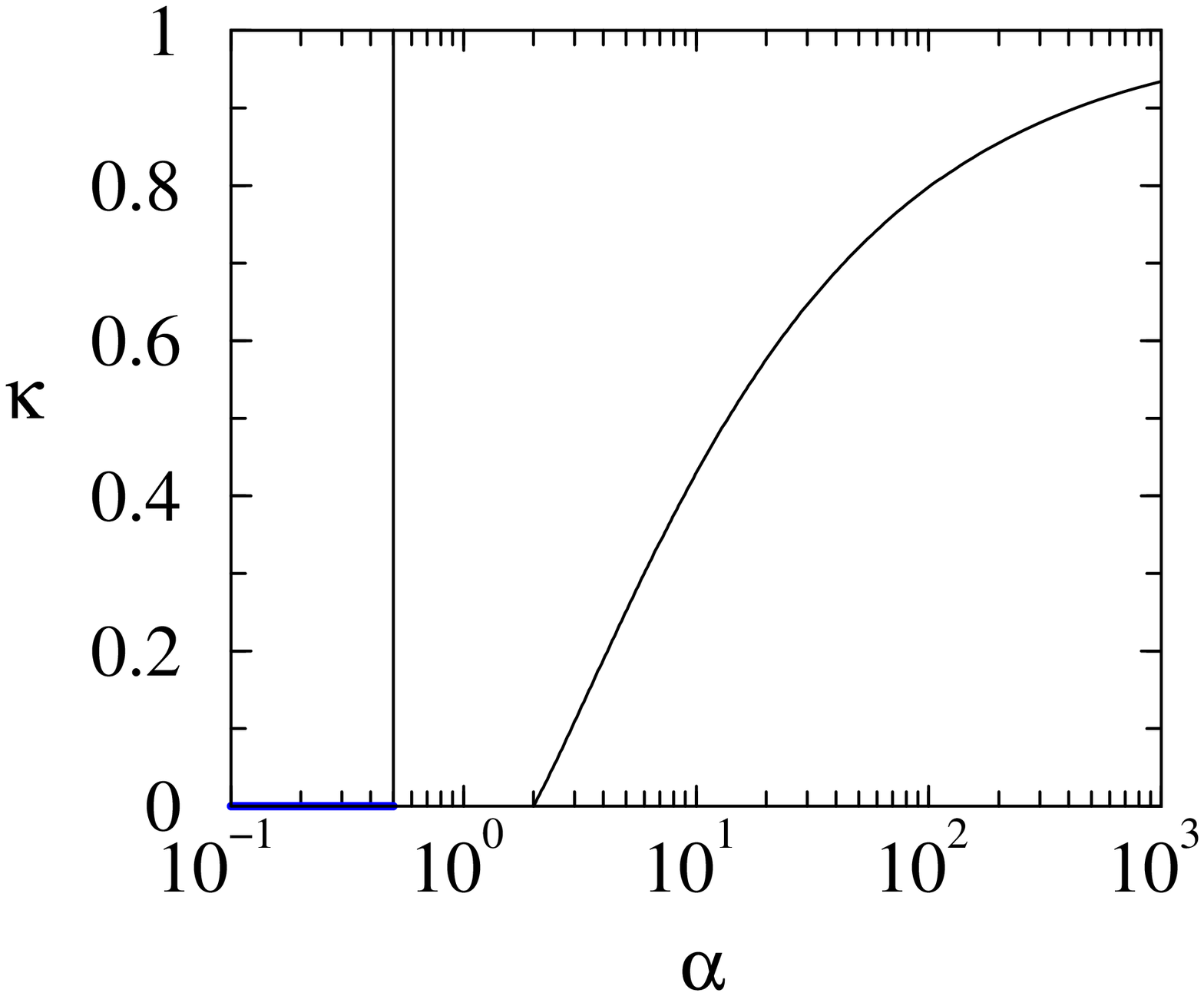}}
%\put(-8,29){\Large $r$} \put( 19,3){\Large $\alpha$}
\put(17,22){\large\bf F} \put(30,15){\large\bf O}
\put(2.5,25){\large\bf AF}
\put(12,35){\vector(-1,0){3}}
\put(13,34.5){$\alpha_{c2}$}
\put(24,30){\vector(1,0){3}}
\put(20,29.5){$\alpha_{c1}$}
\thicklines
\put(-8,13){\vector(2,-1){14}}
\put(-15,13){\large\bf AFO}
\end{picture}
\vspace*{-8mm} \caption{Phase diagram in the $(\alpha,\kappa)$-plane. {\bf O} marks the ergodic {\bf oscillatory} regime at finite integrated response, {\bf F} is the unbounded, but ergodic {\bf frozen} phase, {\bf AF} the phase with {\bf anomalous} response at $\alpha<\alpha_{c2}=1/2$ and $\kappa>0$ in which the system is {\bf frozen}. Finally {\bf AFO} marks the phase with {\bf anomalous} response at $\kappa=0, \alpha<\alpha{c2}=1/2$, in which both {\bf frozen} and {\bf oscillatory} solutions are found depending on initial conditions.} \label{phasediagram}
\end{figure}

The resulting phase diagram is depicted in Fig. \ref{phasediagram}, and we will now turn to a brief discussion of its main features, treating the two cases $\kappa=0$ and $\kappa>0$ separately:

\subsection{$\kappa=0$}
For the game without impact-correction, we find three distinct phases:
(i) an ergodic phase with {\bf oscillatory} behaviour of the
correlation function at $\alpha>\alpha_{c1}=2$ ({\bf O}), (ii) an
unbounded {\bf frozen} (but still ergodic) phase at intermediate
$\alpha\in[1/2,2]$ ({\bf F}), and (iii) a phase with {\bf anomalous}
response below $\alpha=\alpha_{c2}=1/2$ , in which we can find both
{\bf oscillatory} and {\bf frozen} states ({\bf AFO}) , depending on
initial conditions (as illustrated in Fig. \ref{fig:etaeq0}). Note
also that the volatility displays a characteristic minimum at
$\alpha=\alpha_c=1/2$, very much like the fluctuations of the original
MG (we find $\sigma^2(\alpha_c)=0$, however, whereas the volatility in
the conventional MG is always strictly positive at finite values of
$\alpha$). In contrast to the different spherical model discussed in
\cite{GallCoolSher03} we find no discontinuities of the volatility or
oscillation amplitude across any of the transitions.
\subsection{$\kappa>0$}
The behaviour of the model with impact correction is qualitatively similar to the case $\kappa=0$ for large values of $\alpha$ and as long as $\kappa\neq 1$. There we find an oscillatory phase at $\alpha>\alpha_{c1}(\kappa)$ separated from a frozen (but ergodic) phase at intermediate values of $1/2<\alpha<\alpha_{c1}(\kappa)$. The critical value $\alpha_{c1}(\kappa)$ increases as $\kappa$ is increased, with $\lim_{\kappa\to1}\alpha_{c1}(\kappa)=\infty$, see Eq. (\ref{eq:alphac1}), so that no oscillatory phase is found for the game with full impact-correction ($\kappa=1$). Simulations confirm indeed that the game with full impact correction is in fully frozen state ($c_0=1$) for all $\alpha$ , see Fig. \ref{fig:impcorrfig}, in analogy with the conventional MG, where the same feature is observed at $\kappa=1$ \cite{DeMaMa01}. We note however, that the spherical model does not appear to exhibit a simultaneous onset of memory and replica symmetry breaking at finite susceptibility $\chi$, as found in the conventional MG for $0<\kappa<1$ \cite{HeimDeMa01}. Instead we find that anomalous response ($\chi\to\infty$) sets in at $\alpha=1/2$ irrespective of $\kappa$.

\begin{figure}[t]
\vspace*{-26mm} \hspace*{35mm} \setlength{\unitlength}{1.8mm}
\begin{picture}(100,55)
\put(-8,-0){\epsfysize=35\unitlength\epsfbox{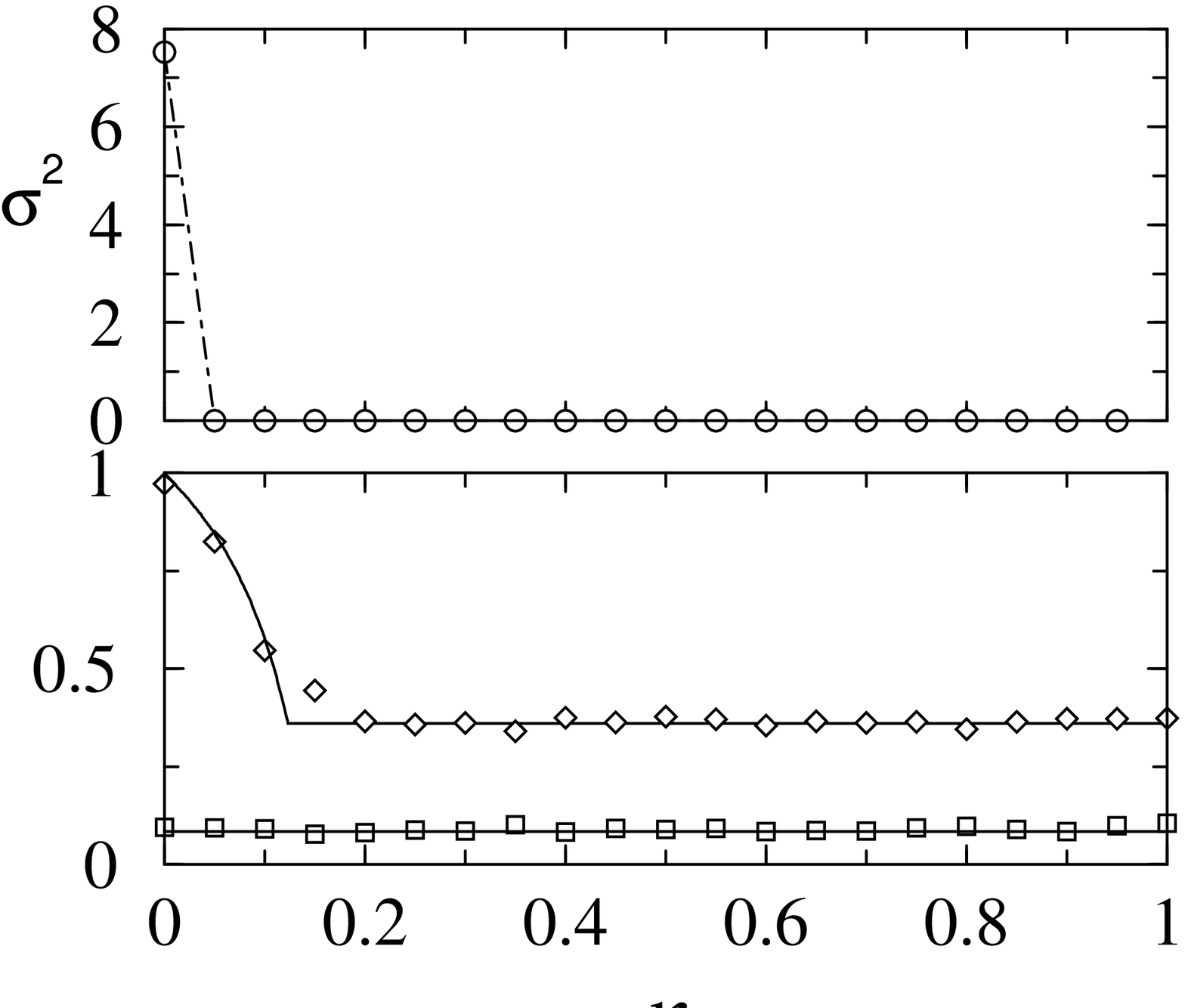}}
\end{picture}
\vspace*{-8mm} \caption{$\sigma^2$ vs $\kappa$ for different values of $\alpha$ at initial bias $q_0=0.01$. Upper panel: $\alpha=0.1$ (circles), where the system is in a non-ergodic state for all $\kappa$. Lower panel: $\alpha=3.16$ (diamonds) and $\alpha=1$ (squares). For $\alpha=3.16$ the system is in the ${\bf F}$ phase above a critical value of $\kappa\approx 0.12$, and in an oscillatory state ${\bf O}$ below $\kappa\approx 0.12$. For $\alpha=1$ the {\bf F} is realised at all $\kappa$. Both panels: markers are from simulations, solid lines from theory for the respective phases. The dashed line in the upper panel is a guide to the eye.}\label{fig:jump}
\end{figure}

Below $\alpha=1/2$ the system is hence in a state of anomalous
response both for $\kappa=0$ and for $\kappa>0$. However, we do not
observe any oscillatory states in simulations in this regime for
$\kappa>0$, while both oscillatory and frozen states are found below
$\alpha=1/2$ in the game without impact correction. In the above phase
diagram we refer to the phase with anomalous response as {\bf AF} for
$\kappa>0$, as opposed to {\bf AFO} for $\kappa=0$. In particular the
volatility $\sigma^2$ turns out to vanish in this {\bf AF} phase at
$\kappa=0$, whereas oscillatory high-volatility solutions can be found
for $\kappa=0, \alpha<1/2$ and small initial bias. This results in a
discontinuity of $\sigma^2$ at $\kappa=0^+$ at any fixed value
$\alpha<1/2$ (and for sufficiently small initial bias), as shown in
the upper panel of Fig. \ref{fig:jump}. We here plot $\sigma^2$ as a
function of $\kappa$ at fixed $\alpha=0.1$ and find that $\sigma^2=0$
for all $\kappa>0$, but that $\sigma^2$ is positive at $\kappa=0$. The
magnitude of this discontinuity depends on $\alpha$ and increases as
$\alpha\to 0$ due to the increasing volatility in the oscillatory
states of the game without impact-correction in this limit (Fig. \ref{fig:etaeq0}). Note that
the volatility is continuous at $\kappa=0^+$ in the ergodic regime
$\alpha>1/2$, see the lower panel of Fig. \ref{fig:jump}. Again very
similar behaviour has been found in the conventional MG and also in
MGs with dilution \cite{MarsChalZecc00, Gall05}.

A further parallel to conventional MGs can be drawn upon noting that
the normalization parameter $\lambda(t)=\lambda_0+\lambda_1 t$ remains
finite asymptotically ($\lambda_1=0$) for all $0\leq \kappa<1$ above
$\alpha_{c1}(\kappa)$ but becomes divergent ($\lambda_1>0$) in the
{\bf F} phase at intermediate $\alpha$. This reflects the behaviour of
the fraction of frozen agents in the conventional MG (corresponding to
runaway solutions) \cite{Book2}, which for $\kappa=0$ is known to rise
sharply as $\alpha$ is increased from the non-ergodic to the ergodic
region. It decreases as $\alpha$ is increased further for conventional
MGs at all $0\leq\kappa<1$ and vanishes in the limit
$\alpha\to\infty$\footnote{Note however that simulations of the game
at $\kappa=0$ in the {\bf AFO} phase indicate that $\lambda_1=0$ even
for high bias. In the conventional game one here finds a large
fraction of frozen agents.}. In addition the fraction of frozen agents
approaches unity as $\alpha\to 0$ for the conventional MG with impact
correction ($\kappa>0$), in analogy with the observed non-zero values
for $\lambda_1$ in the {\bf AF} phase of our model with $\kappa>0$,
see Fig. \ref{fig:impcorrfig}. The spherical model with full impact
correction ($\kappa=1)$ has $\lambda_1>0$ for all $\alpha$, its conventional
counterpart is in a fully frozen state irrespective of $\alpha$.

\section{Replica theory}\label{sec:replica}
We will now turn to the statics of the model. The starting point of the analysis of the statics of the standard MG without spherical constraints is the random function
\be\label{eq:hdef}
\fl ~~ H_\kappa(\boldvarphi)=\frac{1}{\alpha N} \sum_\mu \left(\sum_{i,j}
\xi_i^\mu \xi^\mu_j \varphi_i\varphi_j+2\sqrt{N}\sum_i
\Omega^\mu \xi_i^\mu \varphi_i+N \Omega^\mu\Omega^\mu+\kappa\sum_i \xi_i^\mu\xi_i^\mu(1-\varphi_i^2)\right).
\ee
The computation in the standard MG then proceeds by a minimization of $H_\kappa$ in terms of the variables $\{\varphi_i\}$ using replica techniques. The $\{\varphi_i\}$ here correspond to temporal averages over the (discrete) spins $s_i(t)\in\{-1,1\}$, i.e. one has $\varphi_i=\lim_{t\to\infty}t^{-1}\sum_{\tau\leq t} s_i(\tau)$ so that the $\{\varphi_i\}$ are continuous variables $-1\leq\varphi_i\leq 1$. In the spherical model we have $\sum_i \phi_i(t)^2=N$ at any given time $t$ so that the temporal averages $\varphi_i=\lim_{t\to\infty}t^{-1}\sum_{\tau\leq t} \phi_i(\tau)$ are not individually constrained to the interval $[-1,1]$, but fulfill a global constraint of the form $\sum_i \varphi_i^2\leq N$. That is to say the vector  $\boldvarphi=(\varphi_1,\dots,\varphi_N)$ can take values corresponding to any point {\em within} a sphere of radius $\sqrt{N}$, and is not constrained to its surface. Note that this inequality holds as an equality only if all $\{\phi_i(t)\}$ are time-independent, i.e. if a fully frozen state is reached. To capture cases different from this fully frozen state we first constrain $\boldvarphi$ to a shell of radius $r\sqrt{N}$, and will subsequently minimize the free energy with respect to $0<r\leq 1$. We expect to recover the frozen states as minima of $H$ at $r=1$, and will see below that minima at values $r<1$ correspond to the oscillatory stationary states of the batch game.

At fixed $r$ the replicated partition function at an `annealing temperature'
$T=\beta^{-1}$ is given by
\be
(Z_r)^n=\int_{-\infty}^\infty \left[\prod_{ia} d\varphi_i^a\right] \left[\prod_a\delta\left(r^2N-\sum_i (\varphi_i^a)^2\right)
\exp\left[-\beta H_\kappa(\boldvarphi^a)\right]\right],
\ee
where we have introduced replicated fields
$\{\varphi_i^a\}$. $a=1,\dots, n$ is a replica index, and
$\boldvarphi^a=(\varphi_1^a,\dots,\varphi_N^a)$. The delta-functions
in the above partition function enforce the spherical constraints on
each of the sets $\{\varphi_i^a\}$, $a=1,\dots, n$.

We will be interested in the limit $\beta\to\infty$ of the above
expression eventually, as the minima of $H_\kappa$ are found as
\be
\lim_{N\to\infty} \mbox{min}_{\{\varphi\}}
\frac{H_\kappa(\boldvarphi|r)}{N}=-\lim_{\beta\to\infty}\lim_{N\to\infty}\lim_{n\to
0}\frac{\overline{Z^n}-1}{\beta N n}.
\ee
The notation $H_\kappa(\boldvarphi)|r)$ indicates that the fields are
restricted to radius $r$ at this stage. The computation of
$\overline{(Z_r)^n}$ is lengthy but standard and follows the lines of
\cite{ChalMarsZhan00, Gall05}. Upon making a replica symmetric ansatz
one ends up with the following expression for the free energy given $r$
\BE\label{RSf} 
f_{r,RS}(Q,q,S,s,\rho)&=&\frac{\alpha}{2\beta}\log\left[
1+\frac{\beta}{\alpha}(Q-q)\right]+\frac{\alpha}{2}\frac{(1+q)}{\alpha+\beta(Q-q)}\nonumber\\
&&+\frac{\alpha\beta}{2}(SQ-sq)-\frac{1}{\beta}
\bra\log\int_{-\infty}^\infty d\varphi \exp\left(-\beta
V_z(\varphi)\right)\ket_z\nonumber \\
&& +\frac{\kappa}{2}(1-Q)+\frac{1}{\beta}\rho(r^2-Q),
\EE
where $Q$ and $q$ are the diagonal and off-diagonal elements of the
replica symmetric overlap matrix respectively, and $S$ and $s$ their
conjugate variables. $\rho$ is a Lagrange parameter enforcing the
spherical constraint $Q=r^2$. $V_z(\varphi)$ is an effective potential
defined by
\be
V_z(\varphi)=-\sqrt{\alpha s}z\varphi-\frac{\alpha\beta}{2}(S-s)\varphi^2
\ee
and $\bra\dots\ket_z$ denotes an average over the standard Gaussian variable $z$.

The corresponding saddle-point equations are obtained by
working out the variations of $f_{r,RS}$ with respect to the parameters
$\{Q,q,S,s,\rho\}$. By construction, $\partial f_{r,RS}/\partial\rho=0$ implies $Q=r^2$. The other saddle-point equations are found along the lines of 
\cite{ChalMarsZhan00,Gall05}. In the limit $\beta\to\infty$ we will be looking for solutions
with $\lim_{\beta\to\infty} q=\lim_{\beta\to\infty} Q$ and
$\lim_{\beta\to\infty} s=\lim_{\beta\to\infty} S$, and one defines the finite quantity $\chi=\lim_{\beta\to\infty}\frac{\beta}{\alpha}(Q-q)$. After some algebra the saddle-point equations lead to
%\footnote{We do not report the final result for the Lagrange parameter $\lim_{\beta\to 0}\beta^{-1}\rho$ here, as it does not affect the final expression for the free energy. {\bf to do: check correspondence with dynamics, i.e. with $\lambda$, $\lambda_0$}.}
\be\label{eq:chistat}
\chi=[\sqrt{\alpha}\sqrt{1+1/r^2}-1]^{-1},
\ee
in direct correspondence to (\ref{eq:chiosc}). The free energy at
zero temperature reduces to
\BE\label{eq:fRSofr}
\lim_{\beta\to\infty} f_{r,RS}&=&\frac{1}{2}\frac{1+r^2}{(1+\chi)^2}+\frac{\kappa}{2}(1-r^2)\nonumber\\
&=&\frac{1}{2}\left[\sqrt{1+r^2}-r/\sqrt{\alpha}\right]^2+\frac{\kappa}{2}(1-r^2).
\EE
Finally we minimize this expression with respect to $0<r\leq 1$. If the minimum is taken at $r=1$ we expect the system to be fully frozen, while a minimum at $r<1$ corresponds to an oscillatory state as found in the dynamics.
For $\alpha>\alpha_{c1}$ (with $\alpha_{c1}$ as given in
Eq. (\ref{eq:alphac1})) the absolute minimum of (\ref{eq:fRSofr}) in
the interval $r\in[0,1]$ is identified as
\BE
r^2=\left[\frac{1+(1-\kappa)\alpha+\sqrt{[1+(1-\kappa)\alpha]^2-4\alpha}}{2\sqrt{\alpha}}-1\right]^{-1}<1.
\EE
For $\alpha<\alpha_{c1}$ the minimum is attained at $r=1$.   Hence replica theory leads to the same expressions for the persistent order parameters as the generating functional analysis (see e.g. Eq. (\ref{eq:c0osc})), and in particular we find the same phase boundary between the {\bf O} and {\bf F} phases\footnote{Note that while the static approach allows one to distinguish between frozen phases ($c_0=1$) and phases which are not fully frozen ($c_0<1$), the shape of the correlation function cannot be determined from the replica analysis presented here. The referral to the latter phase as `oscillatory' is therefore motivated by the previous analysis of the dynamics of the batch game}. Furthermore, the frozen phase ($r=1$) obtained from the statics breaks down when $\chi\to\infty$, i.e. at $\alpha=\alpha_{c2}=1/2$ (see Eq. (\ref{eq:chistat})), again in agreement with the phase diagram obtained from the dynamics.

\section{Continuous-time limit and general time-step}\label{sec:conttime}
\subsection{Continuous time}
In this section we will consider a continuous time limit of our model. For simplicity we will restrict the discussion to the case without market impact correction, a generalization to non-zero values
of $\kappa$ is straightforward. We will thus consider
\be\label{eq:conttime}
\frac{d}{dt} q_i(t)=-h_i-\sum_j J_{ij}\phi_j(t),
\ee 
where $\phi_i(t)=q_i(t)/\lambda(t)$, with $\lambda(t)>0$ chosen such
that $\sum_{i=1}^N \phi_i^2(t)=N$ \footnote {Note that this is similar
but not identical to a Langevin equation $\dot q_i(t)=-\frac{\partial
{\cal H}}{\partial q_i}$ with ${\cal H}=\frac{1}{2}\sum_{ij}J_{ij}q_i
q_j+\sum_i h_i q_i$. }. The effective single-agent process corresponding to (\ref{eq:conttime}) reads
\be
\dot q (t)=-\alpha \int dt^\prime
(\id+G)^{-1}(t,t^\prime)\phi(t^\prime)+\theta(t)+\sqrt{\alpha}\eta(t)
\ee
(see also \cite{Crisanti} for $p$-spin spherical spin glass models in
continuous time). The order parameters $C, G$ and $\lambda$ are to be
determined self-consistently according to
\BE
&\bra \eta(t)\eta(t^\prime)\ket_*  =
\left[(\id+G)^{-1} D (\id+G^T)^{-1}\right](t,t^\prime),&\\
&C(t,t^\prime)=\bra \phi(t)\phi(t^\prime)\ket_*, ~~~~~
G(t,t^\prime)=\frac{\delta}{\delta\theta(t^\prime)}\bra
\phi(t)\ket_*,~~~~~C(t,t)=1.&
\EE
Conversion into closed equations for the correlation and response
functions gives
\BE
\frac{\partial}{\partial t}\left[\lambda(t) C(t,t^\prime)\right]&=&-\alpha\left((\id+G)^{-1}C\right)(t,t^\prime)\nonumber
\\
&&+\alpha
\left((\id+G)^{-1}D(\id+G^T)^{-1}G^T\right)(t,t^\prime),\\
\frac{\partial}{\partial t} \left[\lambda(t) G(t,t^\prime)\right]&=&-\alpha\left((\id+G)^{-1}G\right)(t,t^\prime)+\delta(t-t^\prime).
\EE
Assuming that a time-translation invariant bounded state is reached, we may
again write these equations in terms of the Fourier transforms of $C$ and
$G$, and find ($\lambda(t)\equiv\lambda_0$)
\BE
i\omega\lambda_0\widetilde
C(\omega)&=&-\alpha\frac{\widetilde C(\omega)}{1+\widetilde G(\omega)}+\alpha
\frac{\widetilde D(\omega)\widetilde G^*(\omega)}{|1+\widetilde G
(\omega)|^2}\label{eq:langft2}\\
i\omega\lambda_0 \widetilde G(\omega)& = &-\alpha\frac{\widetilde
G(\omega)}{1+\widetilde G(\omega)}+1 \label{eq:langft1}
\EE
Using $D(t-t^\prime)=1+C(t-t^\prime)$ and writing $\chi=\widetilde G(0)$
as usual, we find that the equation for $C$ gives
\be
\left(i\omega|1+\widetilde
G(\omega)|^2+\alpha\right)C(\omega)=\alpha\chi 2\pi\delta(\omega).
\ee
Similarly to calculation in the spherical MG in discrete time, we
conclude that $\widetilde C(\omega)$ can be non-zero only for
$\omega=0$. In contrast to the model in discrete time, there are no
oscillatory solutions. This does not seem to be surprising since the
continuous-time dynamics of (\ref{eq:conttime}) can be expected to
effectively correspond to a system with asynchronous random updating
(as opposed to the parallel dynamics considered earlier) for which
also in the conventional MG model global oscillations are suppressed
\cite{GallSher05}. While this correspondence to asynchronous updating in the MG still deserves further attention we note that synchronous updating in neural networks results in a description of the effective dynamics in terms of a discrete-time process, while asynchronous updating leads to equations in continuous time; see for example \cite{Cool00b,BolleBlan04} for further details.

It is easy to check that assuming a fully frozen state ($c_0=1$) in
Eqs. (\ref{eq:langft2}) and (\ref{eq:langft1}) (evaluated at
$\omega=0$) implies $\alpha=2$, so that we conclude that there can be
no extended phase with bounded values of $\lambda(t)$ in the model in
continuous time, neither a frozen nor an oscillatory one. It would
then appear sensible to revert to our unbounded state
$\lambda(t)=\lambda_0+\lambda_1 t$, with $\lambda_1>0$ and to assume
that $\phi(t)$ in the effective process approaches a fixed point (with
value dependent on the realization of the single-particle noise). This
then leads to Eq. (\ref{eq:vnull}) (where now $v=\lim_{t\to\infty}
\dot q$), and hence to exactly the same solutions for $\chi$ and
$\sigma^2$ as in the {\bf F} phase of the game in discrete time. In
particular one expects a breakdown of ergodicity at $\alpha=1/2$ also
for the model in continuous time.

\subsection{Arbitrary, but finite time-step}
One may ask whether this model in continuous time can be obtained as
the limit $\delta\to 0$ of a model with time-step $\delta$:
\be\label{eq:batchdelta}
\frac{q_i(t+\delta)-q_i(t)}{\delta}=-h_i-\sum_j J_{ij}\phi_j(t).
\ee
$\delta$ can here also be understood as a learning rate, with which
the agents process information and adjust their score updates and
trading actions. We note that the left-hand side can be written as
$[\lambda(t+\delta)\phi_i(t+\delta)-\lambda(t)\phi_i(t)]/\delta$ upon
using $\phi_i(t)=q_i(t)/\lambda(t)$, so that the only effect of the
arbitrary time-step is a re-scaling of the $\{\lambda(t)\}$. Numerical
simulations (not shown here) confirm that the oscillation amplitude of
the correlation function $C(t-\tp)=\bra\phi(t)\phi(\tp)\ket_\star$ in
the stationary state and the volatility are not affected by the
introduction of a general time-step, but that the only effect is a
rescaling of the $\{q_i(t)\}$ relative to the $\{\phi_i(t)\}$. This is
at variance with the spherical model of \cite{GallCoolSher03}, in
which the phase diagram changes as the time-step (or equivalently the
learning rate) is modified, as sketched in the appendix. Note that
a subtle dependence of fluctuations on the learning rate has been
observed in a different agent-based model with continuous degrees of
freedom in \cite{Bergetal,DeMaGall05}.

\section{On-line game}\label{sec:online}
Finally, it is interesting to compare the stationary states of the
on-line model (defined by Eq. (\ref{eq:sphericalonline})) with those
of the batch game. The persistent order parameters (such as $c_0$ and
$\chi$) are known to be identical in the stationary ergodic states of
conventional batch and on-line MGs respectively \cite{Book2}, but
differences in the volatility are in principle to be expected, as the
latter depends not only on the persistent parts of the dynamical order
parameters, but also explicitly on their transients. While these
differences between batch and on-line games are negligible in
conventional MGs with uncorrelated strategy assignments, crucial
differences in their quantitative values and qualitative behaviour
were reported for anti-correlated strategies in
\cite{GallSher05}. These differences can be attributed to persistent
oscillatory behaviour of the batch game.

\begin{figure}[t]
\vspace*{1mm}
\begin{tabular}{cc}
\epsfxsize=60mm  \epsffile{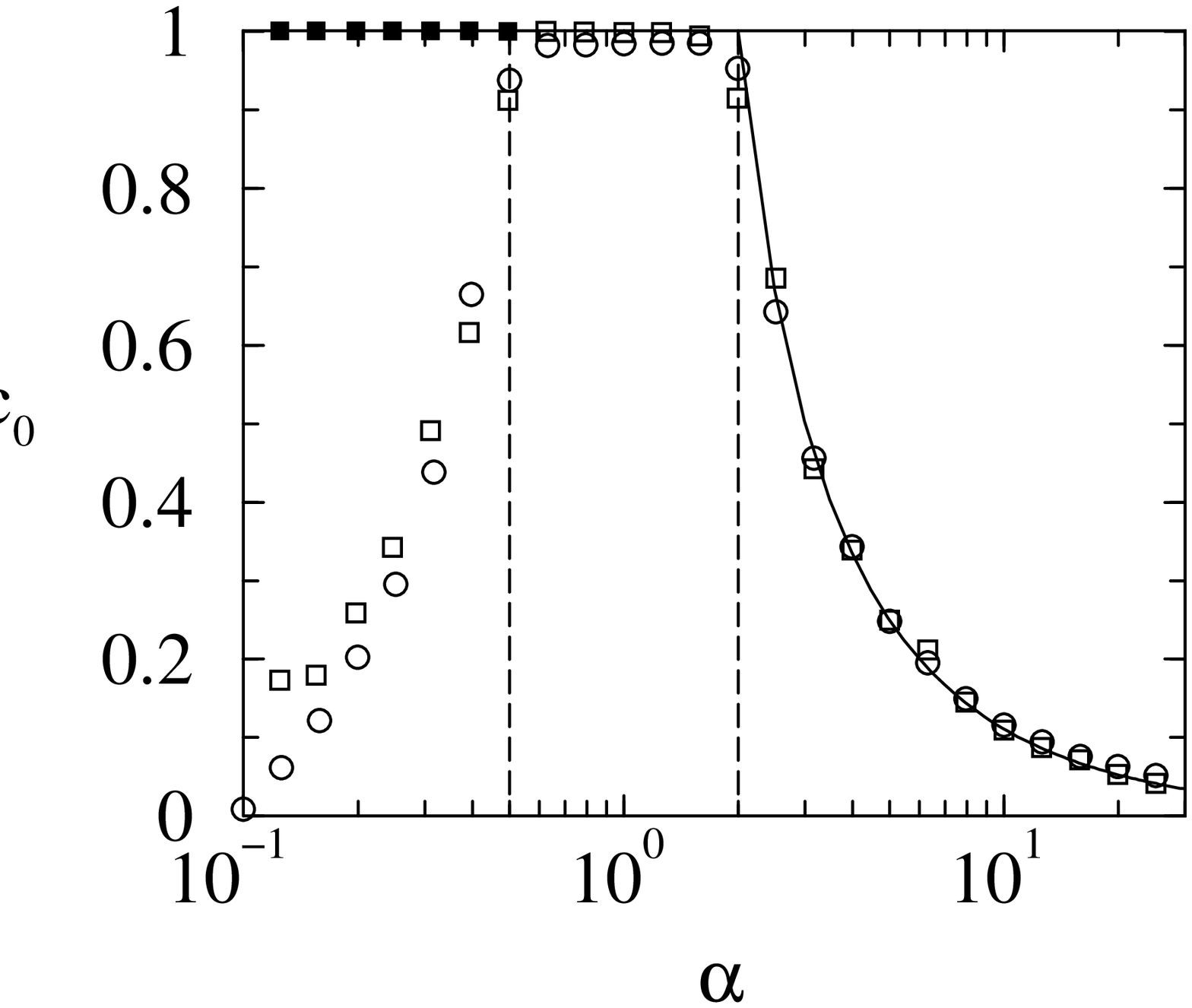} ~~~&~~~
\epsfxsize=60mm  \epsffile{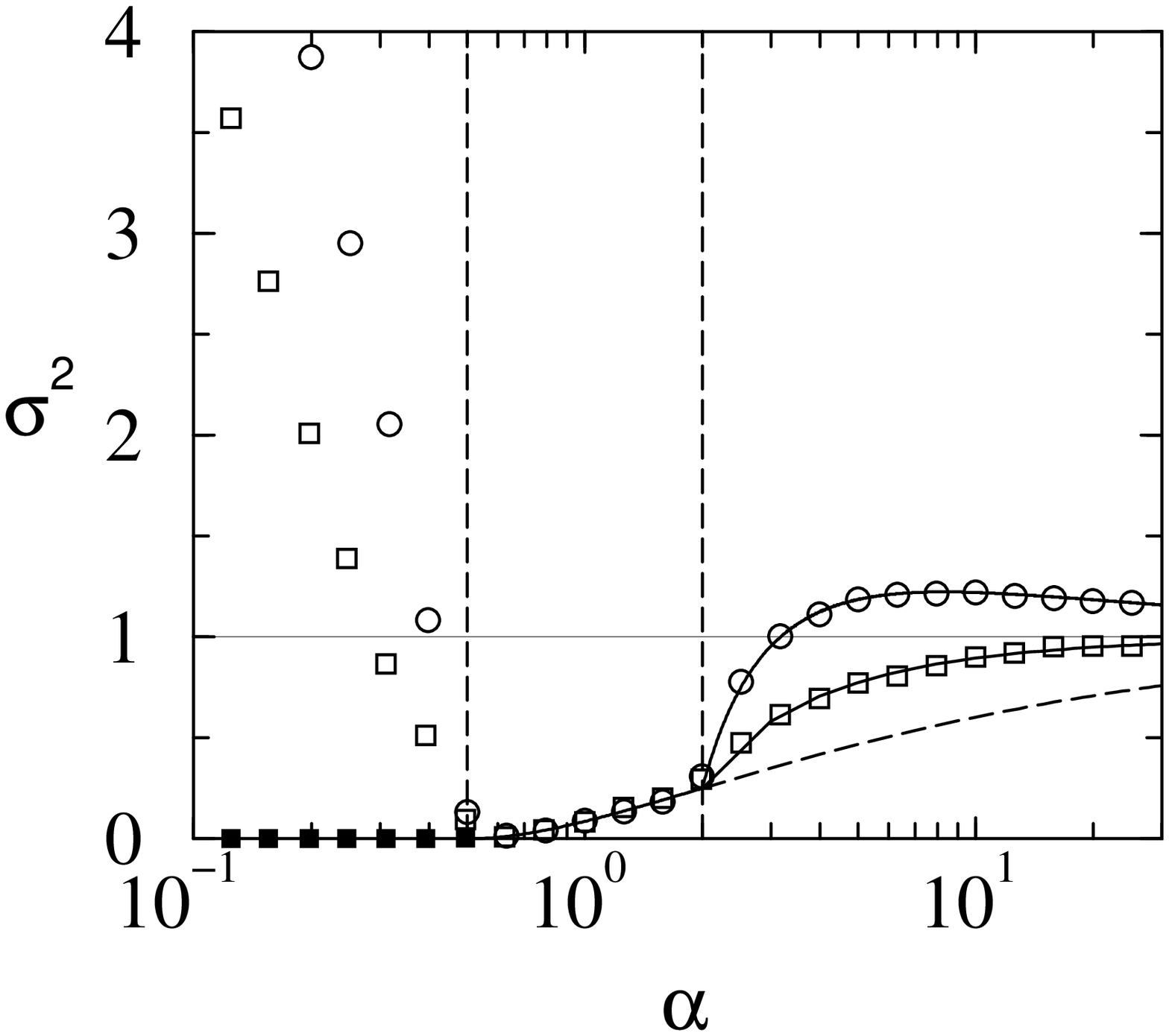} \\ 
\end{tabular}
\vspace*{4mm} \caption{Persistent correlation $c_0$ and volatility $\sigma^2$ for the batch and on-line games without impact correction ($\kappa=0$). Open circles are obtained from simulations of the batch game (simulation parameters as in the other figures), open squares correspond to the on-line game (simulations performed at $\alpha N^2=10^4$, run for $100000$ on-line steps, $20$ samples of the disorder). $q_0=0.01$ in both cases. Solid squares correspond to the on-line game with initial bias $q_0=3$. Solid lines are theoretical predictions, with the upper line for $\sigma^2$ in the high-$\alpha$ phase corresponding to the batch theory (Eq. (\ref{eq:voldef})) and the lower solid line to the on-line approximation (\ref{eq:onlineapprox}). The expression for the fully frozen phase has been continued into the high-$\alpha$ regime (where it is no longer valid neither for the on-line nor the batch game) as a dashed line. \label{fig:online}}
\end{figure}

We compare results from numerical simulations of the batch and on-line
games in Fig. \ref{fig:online}. We restrict the discussion to the case
without market impact correction here ($\kappa=0$). As expected we
observe an ergodic/non-ergodic transition at $\alpha=\alpha_{c2}=1/2$
also in the on-line game and find that the persistent correlation
$c_0$ of the on-line game takes the same values as that of the batch
game in the ergodic regime. In particular $c_0=1$ in both games in the
fully frozen state at intermediate $\alpha$. Within the accuracy of
the simulations the locations of the transitions are insensitive to
the timing of adaptation (on-line versus batch) and in agreement with
the batch theory.

The volatility $\sigma^2$, however, shows a more intricate behaviour,
see right panel of Fig. \ref{fig:online}. While the volatilities of
the two games agree perfectly in the frozen ergodic phase at
intermediate $\alpha\in [1/2,2]$ and are there given by
Eq. (\ref{eq:volF}), one finds differences between the batch and
on-lines cases in the regime of high values of $\alpha$, where the
batch system oscillates in its stationary state. In this phase at
$\alpha>2$ the volatility of the batch game is given by
Eq. (\ref{eq:voldef}), and contains contributions from the persistent
as well as the oscillatory parts of the correlation and response
functions. As shown in Fig. \ref{fig:online}, the volatility of the
on-line game is consistently smaller than that of the batch game in
this regime. The spherical game presented here therefore presents
another example of a MG in which the market fluctuations of the batch
and on-line games differ from each other. As in \cite{GallSher05}
these differences can be attributed to the presence of persistent
oscillations present in the batch game, but not in its on-line
counterpart. Indeed, the volatility of the on-line game is
consistently larger than the theoretical prediction one would obtain
in a fully frozen state, see Fig. \ref{fig:online}, and the now
standard approximation
\be\label{eq:onlineapprox}
\sigma^2=\frac{1}{2}\left[\frac{1+c_0}{(1+\chi)^2}+(1-c_0)\right]
\ee
for the volatility in on-line games, obtained by suitably removing all
non-persistent contributions to $C$ and $G$ (for details see
\cite{CoolHeim01,Book2}) gives near perfect agreement with the numerical
data. While (\ref{eq:onlineapprox}) is at present only an approximate
expression, an exact dynamical theory appears feasible for the present
spherical game also with on-line dynamics (for example along the lines of
\cite{CoolHeim01}) and might allow one to derive an exact result for
the volatility. Finally, we note that started from equal initial
conditions the volatility of the on-line game is consistently smaller
than that of the batch game also for $\alpha<1/2$ (if the initial bias
in this non-ergodic regime is small enough not to end up in a fully
frozen regime, where $\sigma^2=0$ for both games). We recall that
under these circumstances the stationary state of the batch game is
again an oscillatory one.

\section{Concluding remarks}
In summary we have presented a detailed analysis of the stationary
states of a spherical version of the MG, both from a static and a
dynamical point of view. While the present model defines only a minor
modification of an earlier spherical MG, we find that some of its
features bear more similarity with those of the original MG. In
particular one finds a non-ergodic regime, characterized by a strong
dependence of the stationary states on initial conditions, while the
spherical model of \cite{GallCoolSher03} turned out to be ergodic for
all values of the model parameters. Also the model presented here does
not display any discontinuities in the volatility, but the
characteristic minimum of $\sigma^2$ at the point at which the onset
of non-ergodicity occurs. Despite this increased resemblance to the
original MG, the model is still exactly solvable in the ergodic
phases, and in particular exact expressions for the volatility can be
obtained. The model also allows to study the dependence of
volatilities on the timing of adaptation of the agents. Similar to
\cite{GallSher05} one finds that batch games in their oscillatory
phases exhibit larger fluctuations of the total bid than an on-line
game with the same model parameters. We are here able to give an exact
expression for the batch volatility, and a very accurate approximation
for the on-line case, whereas in \cite{GallSher05} analytical results
for the volatility of the batch game with fully anti-correlated
strategies were restricted to upper and lower bounds. Hence, the
present spherical model might serve as a starting point for a more
detailed study of the differences between on-line and batch games,
which are reminiscent of the ones recently observed for example in
\cite{BolleBlan04,BolleBlan05} for sequential and synchronous
updating in other spin systems. Two-cycles, i.e. persistent
oscillations, play a crucial role in those studies similar to what is
observed in our model and in the MG with anti-correlated strategies. A
detailed analysis of an on-line version of the two spherical models
might also help to elucidate the role of transients in the dynamics and
the success of the {\em ad-hoc} removal of non-persistent parts of the
correlation and response functions when deriving an approximate
expression for the volatility. Finally, the dependence on initial
conditions observed in the model discussed here might motivate further
studies of its non-ergodic regime. Ideally, one would like to be able
to compute the persistent order parameters as a function of the
initial bias $q_0$, which would most likely involve a detailed
computation of the transient behaviour. One might hope that work along
these lines could also help to shed some light on the non-ergodic,
turbulent regime of the original MG, a thorough analytical
understanding of which is still awaited.

\section*{Acknowledgements}
 This work was supported by the European Community's Human Potential
 Programme under contract HPRN-CT-2002-00319, STIPCO. TG acknowledges
 the award of a Rhodes Scholarship and support by Balliol College,
 Oxford.  The authors would like to thank R K\"uhn and R Stinchcombe
 for discussions leading to the present work, as well as A C C Coolen
 for earlier collaborative work on spherical Minority Games.

\section*{Appendix: Comparison of the two spherical models}\label{sec:appendix}
In \cite{GallCoolSher03} a spherical limit of the dynamics of the MG was taken upon using the update rule
\be\label{eq:oldspherical}
[1+\lambda(t+1)]q_i(t+1)=q_i(t)-h_i-\sum_j J_{ij}q_j(t),
\ee  
where the $\{\lambda(t)\}$ are chosen to impose a global spherical constraint $\sum_{i=1}^N q_i(t)^2=Nr^2$. A perturbation field $\theta(t)$ may be added to the right-hand-side to generate response functions. The control parameters are given by $\alpha$ and $r$, and in order to be consistent with the spherical constraint initial conditions $\{q_i(t=0)\}$ were chosen such that $\sum_i q_i(0)^2=Nr^2$. In this appendix we will first briefly summarise the main findings of \cite{GallCoolSher03} and then extend the analysis of \cite{GallCoolSher03} by studying the approach to its continuous-time limit and by a brief discussion of the statics of this earlier spherical model.
\subsection*{Phase diagram}
Similar to the analysis in the main part of the paper the generating
functional calculation in \cite{GallCoolSher03} leads to a coupled set
of closed equations for the correlation and response functions (note that here one has $C_{t\tp}=\lim_{N\to\infty} N^{-1}\sum_i \overline{\bra q_i(t)q_i(\tp)\ket}$ and $G_{t\tp}=\lim_{N\to\infty} N^{-1}\sum_i \partial \overline{\bra q_i(t)\ket}/\partial \theta(\tp)$):
\begin{eqnarray}
\left[1+\lambda(t+1)\right]C_{t+1,t^\prime} & = & C_{tt^\prime}
 + \alpha  [(\id+G)^{-1} D
(\id+G^T)^{-1}G^T]_{tt^\prime} \nonumber\\ &&
 -
\alpha [ (\id+G)^{-1}C]_{tt^\prime},
 \label{eq:Cold}
\\
\left[1+\lambda(t+1)\right]G_{t+1,t^\prime} & = & G_{tt^\prime} -
\alpha [(\id+G)^{-1}G]_{t t^\prime} + \delta_{tt^\prime},
\label{eq:Gold}
\EE
\begin{figure}[t]
\vspace*{-6mm} \hspace*{35mm} \setlength{\unitlength}{1.3mm}
\begin{picture}(80,55)
\put(-8,-0){\epsfysize=50\unitlength\epsfbox{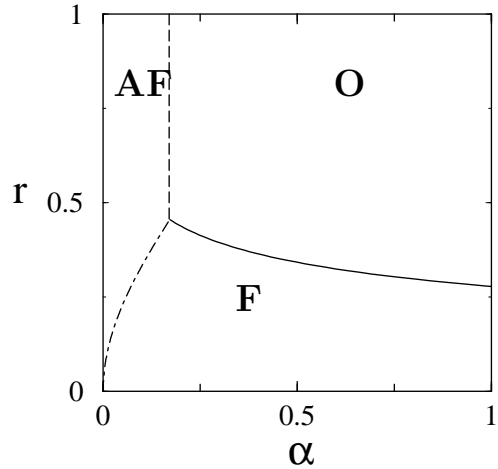}}
%\put(-8,29){\Large $r$} \put( 19,3){\Large $\alpha$}
\put(15,13){\large\bf F} \put(25,35){\large\bf O}
\put(2.5,35){\large\bf AF}
\end{picture}
\vspace*{10mm} \caption{Phase diagram of the spherical MG studied in \cite{GallCoolSher03},
displaying three phases: {\bf O}, oscillating correlation function and
finite integrated response; {\bf F}, frozen phase with finite
integrated response; and {\bf AF}, anomalous frozen phase with
diverging integrated response. Throughout the phase AF the volatility
is zero. The O$\leftrightarrow$F transition and the F$\leftrightarrow$AF transition are
continuous. The discontinuous transition from O$\leftrightarrow$AF occurs at
$\alpha=3-2\sqrt{2}\approx 0.172$. The triple point corresponds to
$\alpha=3-2\sqrt{2}$ and $r=r^*\approx 0.455$. The dotted-dashed line
separating the two frozen phases is given by $\alpha=r^2/(1+r^2)$.}
\label{phasediagram0}
\end{figure}
from which the stationary states of the model can be computed in a
very similar way to the analysis in the main text of this paper. The
resulting phase diagram is depicted in Fig. \ref{phasediagram0}, and
displays three distinct phases {\bf O}, {\bf F} and {\bf AF}, similar
to the model studied in this paper. All phases are bounded however, as
runaway solutions $|q_i(t)|\to\infty$ are impeded by the spherical
constraint on the $\{q_i\}$. While the transitions O$\leftrightarrow$F
and F$\leftrightarrow$AF are continuous, a jump occurs in the
volatility and oscillation amplitude across the O$\leftrightarrow$AF
transition. In particular one finds that the volatility vanishes
identically throughout the ${\bf AF}$ phase, where $\chi=\infty$. The
macroscopic order parameters (including $\sigma^2$) show no dependence
on initial conditions in any part of the phase diagram, in contrast to
the model devised in this paper\footnote{Note that in
\cite{GallCoolSher03} the second moment of the distribution of initial
conditions was fixed by $N^{-1}\sum_i q_i(0)^2=r^2$ for consistency as
this normalization on the $\{q_i(t)\}$ is imposed for all subsequent
times. The freedom of choosing initial conditions in
\cite{GallCoolSher03} was thus limited to distributions with this
fixed value of their second moment which in turn plays the role of a
model parameter. In the present model no constraint on the
$\{q_i(t)\}$ applies so that initial conditions $\{q_i(t=0)\}$ can be
chosen arbitrarily. The analysis in the main part of the paper shows
that different second moments $\lambda(0)^2=N^{-1}\sum_i q_i(0)^2$
lead to different macroscopic stationary states for $\alpha<1/2$.}.
\subsection*{Approach of the continuous-time limit}
The model defined by (\ref{eq:oldspherical}) is easily generalised to arbitrary (but finite) values $\delta$ of the batch time-step:
\begin{eqnarray}
\frac{[1+\lambda(t+\delta)]q_i(t+\delta)-q_i(t)}{\delta}=-h_i-\sum_j J_{ij}~ q_j(t),
\label{eq:sphericalupdatetimestep}
\\[-2mm]
\frac{1}{N}\sum_i q^2_i(t)=r^2~~~~{\rm for~all}~ t,
\label{eq:timestepconstraint}
\end{eqnarray} 
and the analysis of the stationary states is easily adapted
accordingly. In contrast to the model discussed in the main part of the
paper, one here observes interesting effects on the phase diagram. For
$1/2<\delta<2$ we find that the phase diagram and types of transitions
are essentially preserved, but that the triple point moves according
to $\alpha=\left(\sqrt{2/\delta}-1\right)^2$ along the line
$\alpha(r)=r^2/(1+r^2)$ as $\delta$ is varied, and that the numerical
values of the locations of the different transitions change
accordingly, as depicted in the left panel of
Fig. \ref{fig:phasediagram12}. In the case of small $\delta<1/2$ the
discontinuous O$\leftrightarrow$AF transition is no longer present, and the range
of the oscillatory phase moves towards large values of $\alpha$ and
$r$ as $\delta$ is decreased, see the right panel of
Fig. \ref{fig:phasediagram12}. In the limit $\delta\to 0$ the
oscillatory phase will eventually no longer be present, so that the
resulting model in continuous time only displays the ${\bf F}$ and
${\bf AF}$ phases, respectively, separated by a continuous transition
at $\alpha=r^2/(1+r^2)$. As before all order parameters can be
computed exactly for arbitrary values of $\delta$ in all phases, and
numerical simulations (not shown here) confirm these results as well
as the changes in the phase diagram \cite{Gallathesis}.
\begin{figure}[t]
\vspace*{-8mm} \hspace*{25mm} \setlength{\unitlength}{1.1mm}
\begin{tabular}{cc}
\begin{picture}(100,55)
\put(-8,5){\epsfysize=50\unitlength\epsfbox{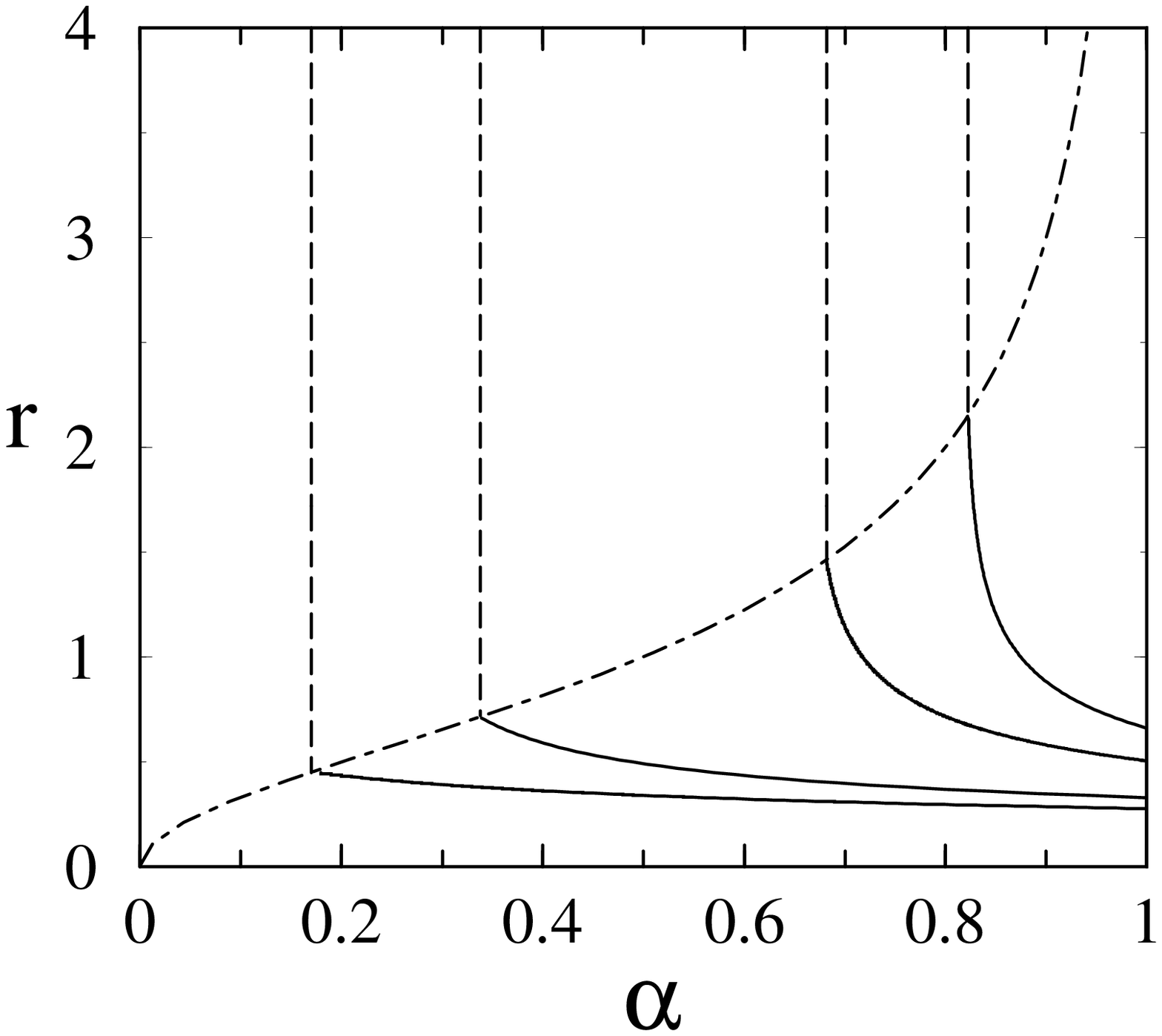}}
%\put(-8,29){\Large $r$} \put( 19,3){\Large $\alpha$}
\put(3,11){\bf F} 
\put(0,30){\bf AF}
\put(40,23){\bf O}
%\put(2.5,35){\large\bf AF}
\end{picture} & 
\begin{picture}(100,55)
\put(-43,5){\epsfysize=50\unitlength\epsfbox{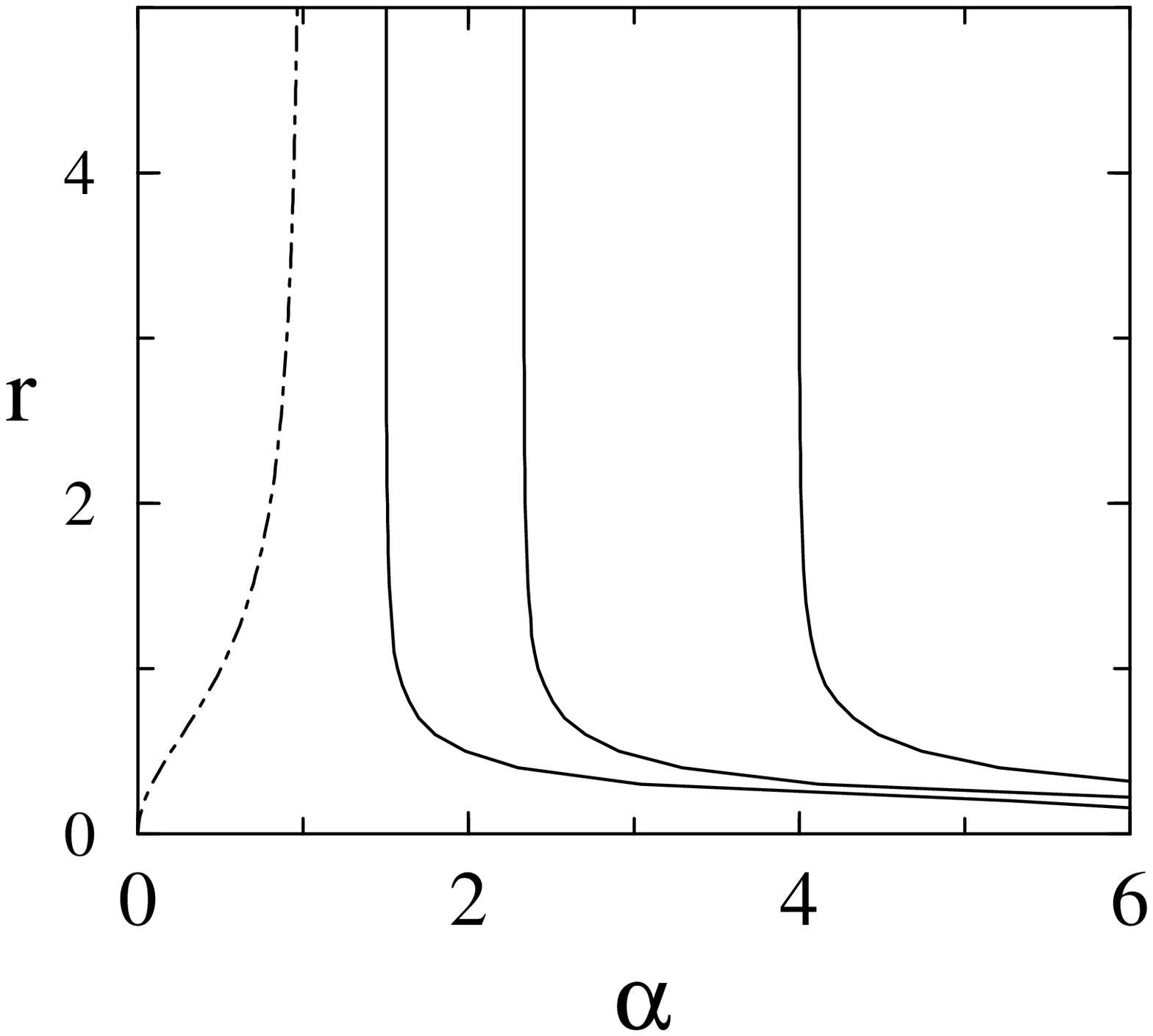}}
%\put(-8,29){\Large $r$} \put( 19,3){\Large $\alpha$}
\put(-32,14){\large\bf F} 
\put(-37,30){\large\bf AF}
\put(2,20){\large\bf O}
%\put(2.5,35){\large\bf AF}
\end{picture}
\end{tabular}
\vspace*{-8mm}\caption{Phase diagram of the model of \cite{GallCoolSher03} for general time-step $\delta$. Left ($\delta>1/2$):  solid lines: continuous O$\leftrightarrow$F transition for $\delta=1.0,0.8,0.6,0.55$ (from left to
right). The dotted-dashed line is given by $\alpha=r^2/(1+r^2)$
and marks the F$\leftrightarrow$AF transition. Right: phase diagram for
$\delta<1/2$: solid lines: continuous O$\leftrightarrow$F transition for
$\delta=0.4,0.3,0.2$ (from left to right). The dotted-dashed line is
given by $\alpha=r^2/(1+r^2)$ and marks the F$\leftrightarrow$AF
transition. } \label{fig:phasediagram12}
\end{figure}
\subsubsection*{Statics}
Finally, we note that the statics of the model (\ref{eq:oldspherical})
can be studied upon minimizing the Hamiltonian $H_{\kappa=0}$ defined
in (\ref{eq:hdef}) \cite{Gallathesis}. Again the variables
$\{\varphi_i\}$ are to be constrained to a sphere of radius $r$, where
now $r$ is a model parameter and fixed (i.e. the free energy is not
minimized with respect to $r$). Indeed the dynamical approach via
generating functionals and the static calculation via replica methods
turn out to deliver the same order parameters in the {\bf F} phase of
the model defined by (\ref{eq:oldspherical}). However, a 
replica approach along the lines of the main part of this paper (in
which ground state solutions are not constrained to the surface of the
sphere with radius $r$ but can lie anywhere within) fails to
reproduce the dynamical results obtained for the oscillatory regime of
(\ref{eq:oldspherical}). It appears that here a more sophisticated
approach based on (pseudo-) Hamiltonians of the Peretto type
\cite{Pere84,SkanCool00} might be required, possibly along the lines
of \cite{BolleBlan04, BolleBlan05}.
\end{document}